\documentclass[aps,a4paper,superscriptaddress,nofootinbib,twocolumn]{revtex4-1}
\usepackage{amsmath,amssymb,gensymb,graphicx,multirow,dcolumn,bm,latexsym,soul,nicefrac,acronym}
\usepackage[mathscr]{euscript}
\usepackage{appendix}
\usepackage[colorlinks,linkcolor=blue,citecolor=blue,urlcolor=blue ]{hyperref}
\usepackage{float}
\usepackage{soul}
\usepackage{verbatim}
\usepackage{color}
\usepackage{acronym}
\usepackage{subfigure}
\usepackage{appendix}
\usepackage{longtable}
\usepackage{footnote}
\usepackage{cancel}
\usepackage{ulem} 

\newcommand{\micro}{\si{\micro}}

\newcommand{\be}{\begin{equation}}
\newcommand{\ee}{\end{equation}}
\newcommand{\bea}{\begin{eqnarray}}
\newcommand{\eea}{\end{eqnarray}}
\newcommand{\bw}{\begin{widetext}}
\newcommand{\ew}{\end{widetext}}
\newcommand{\nn}{\nonumber}

\newcommand{\eq}[1]{Eq.~(\ref{#1})}
\newcommand{\fig}[1]{Fig.~\ref{#1}}
\newcommand{\tab}[1]{Table.~\ref{#1}}

\newcommand{\TRC}{MOE Key Laboratory of TianQin Mission, TianQin Research Center for Gravitational Physics $\&$  School of Physics and Astronomy, Frontiers Science Center for TianQin, CNSA Research Center for Gravitational Waves, Sun Yat-sen University (Zhuhai Campus), Zhuhai 519082, China}

\begin{document}
\title{Sensitivity to anisotropic gravitational-wave background with space-borne detector networks}
\author{Zheng-Cheng Liang}
\affiliation{\TRC}
\author{Zhi-Yuan Li}
\affiliation{\TRC}
\author{En-Kun Li}
\affiliation{\TRC}
\author{Jian-dong Zhang}
\email{Corresponding author: zhangjd9@sysu.edu.cn}
\affiliation{\TRC}
\author{Yi-Ming Hu}
\email{Corresponding author: huyiming@sysu.edu.cn}
\affiliation{\TRC}

\date{\today}

\begin{abstract}
Single gravitational-wave detectors face inherent limitations in detecting the anisotropy of the stochastic background. 
In this work, we explore the sensitivity to anisotropic backgrounds with a network of space-borne detectors. 
We find that the separation between detectors plays an important role in determining the sensitivity. 
For the first time, we observe as large as three orders of magnitude enhancement in detection sensitivity for the multipoles with $l=5$ and 6, compared to coinciding detectors. 
Coordinating and optimizing the separation between two space-borne detectors can significantly enhance the network's sensitivity to the multipole components of the stochastic background. 
For the TianQin + LISA network, benefiting from detector separation, it is possible to achieve sensitivity levels of 2-3 orders of magnitude better than using TianQin or LISA detector alone. 
These findings pave the way to uncover the underlying physics of anisotropy through gravitational-wave detections.

\end{abstract}



\maketitle
\acrodef{SGWB}{stochastic \ac{GW} background}
\acrodef{GW}{gravitational-wave}
\acrodef{CBC}{compact binary coalescence}
\acrodef{MBHB}{supermassive black hole binary}
\acrodef{SBBH}{stellar-mass binary black hole}
\acrodef{EMRI}{extreme-mass-ratio inspiral}
\acrodef{DWD}{double white dwarf}
\acrodef{BH}{black hole}
\acrodef{NS}{neutron star}
\acrodef{BNS}{binary neutron star}
\acrodef{LIGO}{Laser Interferometer Gravitational-Wave Observatory}
\acrodef{LISA}{Laser Interferometer Space Antenna}
\acrodef{TQ}{TianQin}
\acrodef{KAGRA}{Kamioka Gravitational Wave Detector}
\acrodef{ET}{Einstein telescope}
\acrodef{DECIGO}{DECi-hertz interferometry GravitationalWave Observatory}
\acrodef{CE}{Cosmic Explorer}
\acrodef{NANOGrav}{The North American Nanohertz Observatory for Gravitational Waves}
\acrodef{LHS}{left-hand side}
\acrodef{RHS}{right-hand side}
\acrodef{ORF}{overlap reduction function}
\acrodef{ASD}{amplitude spectral density}
\acrodef{PSD}{power spectral density}
\acrodef{SNR}{signal-to-noise ratio}
\acrodef{TDI}{time delay interferometry}
\acrodef{PIS}{peak-integrated sensitivity}
\acrodef{PLIS}{power-law integrated sensitivity}
\acrodef{GR}{general relativity}
\acrodef{PBH}{primordial black hole}
\acrodef{SSB}{solar system baryo}
\acrodef{PT}{phase transition}
\acrodef{PTA}{Pulsar Timing Arrays}
\acrodef{PI}{power-law integrated}
\acrodef{SVD}{singular value decomposition}
\acrodef{FIM}{Fisher information matrix}

\section{Introduction}
The detection of \acp{GW} has provided a new way to observe the Universe, with almost a hundred \ac{GW} events announced so far~\cite{LIGOScientific:2021djp}. 
However, not all \acp{GW} can be directly detected, leading to the existence of a large number of independent and unresolved \acp{GW} that form a \ac{SGWB}~\cite{Christensen:1992wi,Flanagan:1993ix,Allen:1997ad}. 
In the nano-Hertz ($\rm nHz$) band, a number of collaborations have announced the successful detection of \ac{SGWB}~\cite{NANOGrav:2023gor,Xu:2023wog,EPTA:2023fyk,Reardon:2023gzh}. 
Furthermore, there is great potential for future space-borne detectors~\cite{Seto:2020mfd,Liang:2021bde,Fan:2022wio} and ground-based detectors~\cite{LIGOScientific:2017zlf,Chen:2018rzo,LIGOScientific:2019vic} to contribute to the \ac{SGWB} detection. 
An interesting aspect of the \ac{SGWB} is its potential anisotropy, where the intensity exhibits spatial variations~\cite{Hils:1990vc,Geller:2018mwu,Wang:2021djr,Li:2021iva,Li:2023qua,Profumo:2023ybp,Cui:2023dlo}. 
Among the various possible anisotropic \acp{SGWB}, the most promising candidate in the milli-Hertz ($\rm mHz$) band is the foreground generated by the Galactic \ac{DWD}~\cite{Robson:2018ifk,Huang:2020rjf}. 

Space-borne detectors are designed to detect \acp{GW} within the $\rm mHz$ band~\cite{TianQin:2015yph,LISA:2017pwj,Hu:2017mde}, where laser noise is the dominant noise needs to be suppressed by orders of magnitude to enable the successful \ac{GW} detection~\cite{1996OptCo.123..669G}. 
To achieve this goal, \ac{TDI} combination is proposed~\cite{Tinto:1999yr,Tinto:2020fcc}, and among all the \ac{TDI} combinations, the orthogonal channel group $\rm AET$ is commonly used~\cite{Prince:2002hp}. 
The $\rm A/E$ channels in this channel group are expected to detect different polarization of the \ac{GW}, through which the \ac{GW} can be fully characterized. 
The $\rm T$ channel, in contrast, is much less sensitive to the \ac{GW} at low frequencies than the $\rm A/E$ channels~\cite{Hogan:2001jn}. 

After the effective suppression of laser noise, secondary noises, such as optical-path noise and acceleration noise, become prominent in the detector channel~\cite{Krolak:2004xp}. 
To extract the \ac{SGWB} signal from these noises, correlation analysis using either null-channel method~\cite{Tinto:2001ii,Hogan:2001jn,Cheng:2022vct} or cross-correlation method~\cite{Christensen:1992wi,Flanagan:1993ix} is required. 
The null-channel method involves using a signal-insensitive channel to monitor the secondary noises of other channels, while the cross-correlation method is applicable when there is no correlation between the noise of the adopted channels. 
By performing auto- or cross-correlation of the channel output, the correlation measurement of \ac{SGWB} signal can be extracted, which is related to the statistical property, or, equivalently, the \ac{PSD} of \ac{SGWB}~\cite{Allen:1996vm}. 
Additionally, the correlation measurement also includes the antenna pattern, which is determined by the channel geometry and the separation between detectors~\cite{Cornish:2001hg}. 
Accumulating the antenna pattern for all spatial orientations allows the determination of the \ac{ORF}. 
The \ac{ORF} represents the reduction of the correlation measurement relative to the \ac{PSD} of the \ac{SGWB}.

To further investigate the anisotropy of the \ac{SGWB}, it is possible to analyze the antenna pattern by decomposing it into spherical harmonics with multipole coefficients, which are directly related to the multipole moments of the anisotropic \ac{SGWB}~\cite{Allen:1996gp,Cusin:2017fwz}. 
The analysis of the multipole coefficients allows for assessing anisotropy detection from two perspectives. Firstly, 
by calculating the \ac{ORF} corresponding to each multipole coefficient and taking into account the noise level of channel, the detection sensitivity to the anisotropic \ac{SGWB} can be determined. 
This curve helps determine the minimum detectable intensity of the multipole moment within detectors. 
Secondly, one can estimate the values of the multipole coefficients based on data. One method is to compute the frequentist maximum-likelihood estimators of these coefficients~\cite{Mitra:2007mc,Thrane:2009fp,Renzini:2018vkx,Suresh:2020khz,Contaldi:2020rht,Renzini:2021iim,Xiao:2022uvq,Agarwal:2023lzz}. Another method is to utilize Bayesian inference to construct their posterior distributions given prior probability distributions for the signal and noise parameters~\cite{Taylor:2020zpk,Payne:2020pmc,Banagiri:2021ovv,Tsukada:2022nsu,Chung:2023rpq}. These methods facilitate the creation of a ``clean map" depicting the true distribution of \ac{GW} power across the sky~\cite{Mitra:2007mc,Thrane:2009fp,Gair:2014rwa}.

Previous studies have highlighted the limitations of a single space-borne detector in detecting multipole moments beyond the monopole ($l=0$), quadrupole ($l=2$), and hexadecapole ($l=4$) ~\cite{Cornish:2001hg,Kudoh:2004he,LISACosmologyWorkingGroup:2022kbp}, necessitating a loud \ac{SGWB} signal to estimate other multipole coefficients~\cite{Contaldi:2020rht,Banagiri:2021ovv}. 
In comparison to single detectors, the use of a detector network introduces two additional factors, namely the detector plane angle and the detector separation, which have significant influence on the \ac{SGWB} detection. 
For example, for an isotropic background ($l=0$), a pair of detectors with smaller plane angles and shorter separations would be associated with better sensitivities~\cite{Cornish:2001bb,Liang:2022ufy}. 
Extending beyond isotropic background, multipole moments with $l=1$ and $3$ experience a substantial enhancement in \ac{SNR} when using a detector network compared to a single detector~\cite{Ungarelli:2001xu}.

In this paper, we focus on investigating the sensitivity of space-borne detector networks to anisotropy. Through a detailed analysis, we quantitatively examine the factors influencing anisotropy sensitivity. 
We derive a universal correlation of the detector separation on the sensitivity to the anisotropy of the \ac{SGWB}. Our findings indicate that increasing the detector separation can greatly enhance the detection of specific multipoles within the \ac{SGWB}. Crucially, as the order of the multipole moment increases, the benefits derived from larger detector separations become even more pronounced. 
Building upon that premise, we demonstrate the detection sensitivity to the anisotropic \ac{SGWB} through realistic examples. 
We consider two space missions, TianQin and \ac{LISA}, as potential options for the detector network: the TianQin I+II network and the TianQin + LISA network. 
In the TianQin I+II network, the first detector, TianQin, will have a fixed pointing direction, while the second detector, TianQin II, will maintain a perpendicular orientation to TianQin~\cite{Ye:2019txh}. 
On the other hand, the TianQin + LISA network will involve LISA changing the pointing direction over time. 
To assess the detection sensitivity, we utilize the \ac{PLIS} curve~\cite{Thrane:2013oya}. 
This specific sensitivity curve allows us to estimate the minimum detectable energy spectrum density of the \ac{SGWB}. 

This paper is structured as follows. 
In Sec.~\ref{sec:Fundamental}, we review the fundamental properties of stochastic gravitational-wave background. 
We then discuss the analysis for detecting the anisotropy of the \ac{SGWB} in Sec.~\ref{sec:analysis}. 
We analyze the improvement in detection sensitivity to anisotropy with detector separation in Sec.~\ref{sec:result_I}. 
Furthermore, in Sec.~\ref{sec:result_II}, we conduct the realistic case study involving TianQin and LISA. 
Sec.~\ref{sec:conclusion} is devoted to a conclusion and discussion. 

\section{Theoretical Fundamental}\label{sec:Fundamental}
In order to characterize the intensity of \ac{SGWB}, we begin with the metric perturbation in the transverse-traceless gauge. 
Since the \ac{SGWB} is the superposition of \acp{GW}, the metric perturbation $h(t,\vec{x})$ can be expressed as follows:
\bea
\label{eq:h_ab}
\nn
h(t,\vec{x})&=&\sum_{P=+,\times}\int_{-\infty}^{\infty}{\rm d}f\int_{S^{2}}{\rm d}\hat{\Omega}_{\hat{k}}\,
\widetilde{h}_{P}(f,\hat{k})\textbf{e}^{P}(\hat{k})\\
&&\times e^{{\rm i}2\pi f[t-\hat{k}\cdot\vec{x}(t)/c]},
\eea
where $\hat{k}$ is the wave vector of the \ac{GW}, the polarization tensor $\textbf{e}^{P}(\hat{k})$ refers to the polarization $P$ of the \ac{GW}, $c$ denotes the light speed. 
In this work, we focus on the Gaussian-stationary and unpolarized \ac{SGWB}, which allows us to characterize the statistical properties of \ac{SGWB} through the mean value and variance of the Fourier amplitude $\widetilde{h}_{P}(f,\hat{k})$. 
Based on the fact that \ac{SGWB} is composed of a large number of \acp{GW}, it is reasonable to assume that the \ac{SGWB} has a zero-mean Fourier amplitude $\widetilde{h}_{P}(f,\hat{k})$:
\be
\langle\widetilde{h}_{P}(f,\hat{k})\rangle=0.
\ee
On the other hand, the variance of $\widetilde{h}_{P}(f,\hat{k})$ is defined as the \ac{PSD} $\mathscr{P}_{\rm h}$ of \ac{SGWB}:
\be
\label{eq:Ph}
\langle\widetilde{h}_{P}(f,\hat{k})\widetilde{h}^{*}_{P'}(f',\hat{k}')\rangle
=\frac{1}{4}\delta(f-f')\delta_{PP'}\delta^{2}(\hat{k}-\hat{k}')\mathscr{P}_{\rm h}(|f|,\hat{k}),
\ee
where the factor $1/4$ comes from the definition of one-side \ac{PSD} and the average of polarization. 

In the case of an anisotropic \ac{SGWB}, where the \ac{PSD} exhibits a directional preference, it is convenient to expand the \ac{PSD} in terms of spherical harmonics. 
By assuming that the intensity of the \ac{SGWB} has no dependence between direction and frequency, one can focus on characterizing the direction dependence of the \ac{SGWB} through~\cite{Allen:1996gp}
\be
\label{eq:P_h}
\mathscr{P}_{\rm h}(f,\hat{k})
=\sum_{l=0}^{\infty}\sum_{m=-l}^{l}p_{lm}(f)Y_{lm}(\hat{k}),
\ee
with the multipole moment
\be
\label{eq:p_lm}
p_{lm}(f)
=\int_{S^{2}}{\rm d}\hat{\Omega}_{\hat{k}}
\mathscr{P}_{\rm h}(f,\hat{k}) Y^{*}_{lm}(\hat{k}).
\ee

To obtain the {\it all-sky} \ac{PSD} of \ac{SGWB}, the $\mathscr{P}_{\rm h}$ needs to be integrated over all directions: 
\bea
\nn
\label{eq:S2P}
S_{\rm h}(f)&=&
\int_{S^{2}}{\rm d}\hat{\Omega}_{\hat{k}}
\mathscr{P}_{\rm h}(f,\hat{k})\\
&=&
\sqrt{4\pi}\,p_{00}(f).
\eea
\eq{eq:S2P} highlights that the all-sky \ac{PSD} solely captures the monopole moment $p_{00}$, while the higher-order multipole moments associated with the anisotropy remain hidden. 
To study the anisotropy of \ac{SGWB}, it is crucial to establish an estimator that incorporates the quadratic form of $\mathscr{P}_{\rm h}$: 
\bea
\label{eq:A_l}
\nn
\int_{S^{2}}{\rm d}\hat{\Omega}_{\hat{k}}
|\mathscr{P}_{\rm h}(f,\hat{k})|^2
&=&\sum_{l=0}^{\infty}\sum_{m=-l}^{l}p_{lm}(f)p^{*}_{lm}(f)\\
&=&\sum_{l=0}^{\infty}(2l+1)\mathscr{A}_{l}(f),
\eea
where the angular \ac{PSD} plays a significant role in providing insights into the anisotropy across various scales~\cite{Thrane:2009fp}:
\be
\mathscr{A}_{l}(f)=
\frac{1}{2l+1}\sum_{m=-l}^{l}p_{lm}(f)p^{*}_{lm}(f).
\ee
Especially, for an isotropic \ac{SGWB}, 
\be
\mathscr{A}_{0}(f) = \frac{S^{2}_{\rm h}(f)}{4\pi}.
\ee

In addition, to characterize the distribution of energy across different frequencies in \ac{SGWB}, the dimensionless energy spectrum density $\Omega_{\rm gw}$ is commonly used. 
This quantity allows us to specify the ratio of the \ac{GW} energy density ${\rm d}\rho_{\rm gw}$ within the frequency range [$f$,$f+{\rm d}f$] to the critical energy density $\rho_{\rm gw}$~\cite{Allen:1996vm}:
\be
\label{eq:omega_gw}
\Omega_{\rm gw}(f)=\frac{1}{\rho_{\rm c}}\frac{{\rm d}\rho_{\rm gw}}{{\rm d}(\ln{f})},
\ee
where the critical energy density $\rho_{\rm c}=3H_{0}^{2}c^{2}/(8\pi G)$, with the gravitational constant $G$, and the Hubble constant $H_{0}$. 
The relationship between $\Omega_{\rm gw}$ and $S_{\rm h}$ can be established through~\cite{Thrane:2013oya}
\be
\label{eq:Omega2Sh}
\Omega_{\rm gw}(f)
=\frac{2\pi^{2}}{3H_{0}^{2}}f^{3}S_{\rm h}(f).
\ee

\begin{figure}[htbp]
	\centering
    \includegraphics[width=\linewidth]{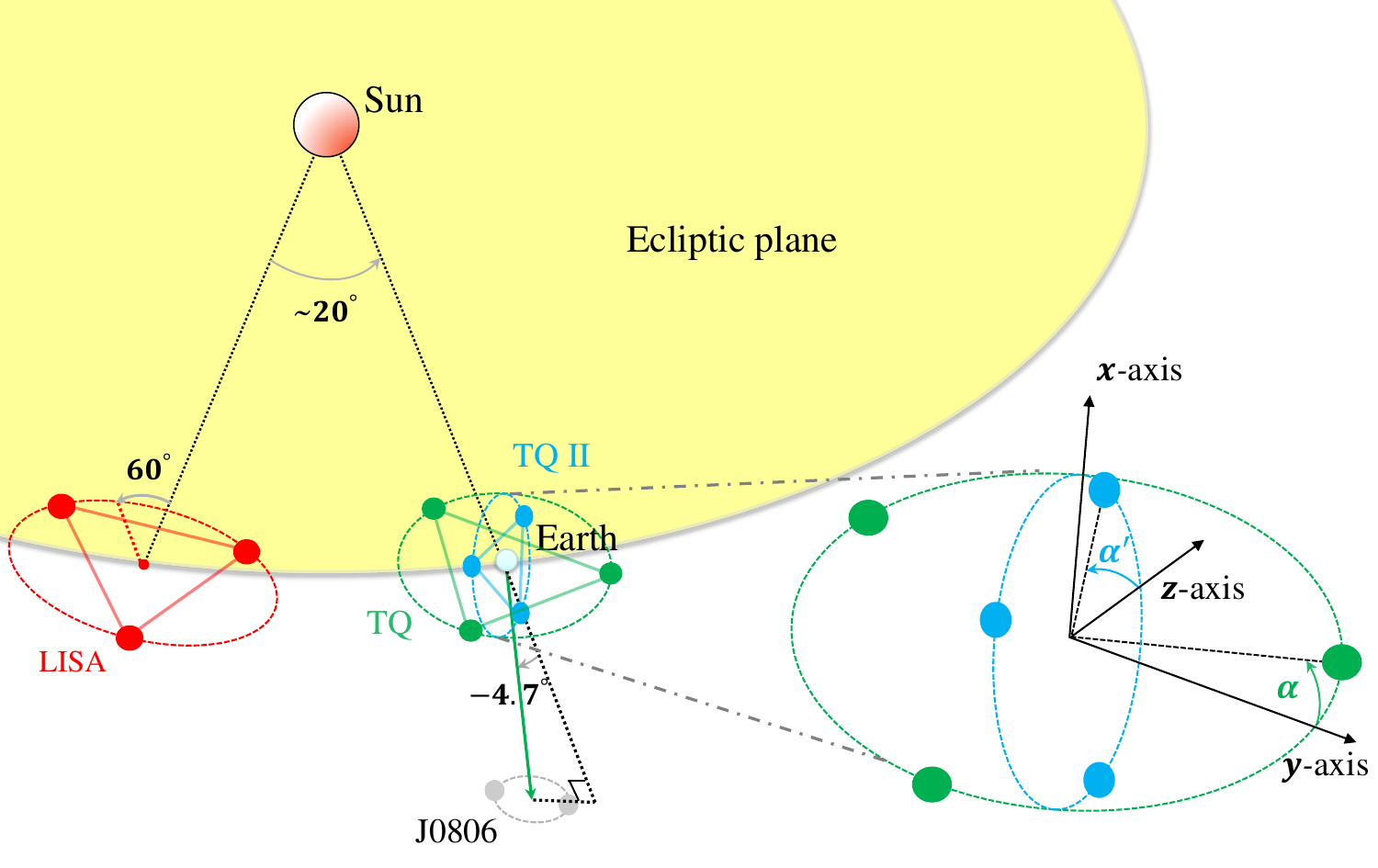}
	\caption{Schematic diagram of TianQin and LISA mission in the ecliptic coordinate.}
	\label{fig:Configuration}
\end{figure}
\section{Detection analysis}\label{sec:analysis}
\subsection{Detector design and channel}
Currently, there are several proposed space missions for detecting \acp{GW}, including LISA~\cite{LISA:2017pwj}, TianQin~\cite{TianQin:2015yph}, Taiji~\cite{Hu:2017mde}, etc. 
In this paper, we constrain our focus on TianQin and LISA. 

As shown in \fig{fig:Configuration}, TianQin is a space-borne \ac{GW} detector comprising three identical satellites orbiting the Earth. 
These satellites have an orbital period of $3.64$ days and are positioned at a radius of approximately $10^{5}\,\, {\rm km}$. 
When in operation, the three satellites will form an equilateral triangle with an armlength $L_{\rm TQ}$ of about $1.7\times10^{5}\,\, {\rm km}$. 
The working mode of TianQin follows a ``three months on + three months off" pattern. 
To ensure continuous detection, a second mission called TianQin II is proposed to bridge the detection gap~\cite{Ye:2019txh}. 
The orbital planes of TianQin and TianQin II are designed to be perpendicular to each other, forming the TianQin I+II network. 
In fact, the initial working mode does not allow for overlapping operating time between TianQin and TianQin II. 
To address this issue, we adopt the improved working mode ``four months on + two months off"~\cite{Ye:2020tze}, which allows for a four-month overlap between TianQin and TianQin II within a year. It is crucial to account for the time lag between the emissions of TianQin and TianQin II, which results in a disparity in their respective initial phase angles within their planes. To simplify subsequent analysis, we introduce the following notation: $\alpha$ and $\alpha'$ represent the initial phase angles of both detectors, while $\gamma_{0}$ is defined as the phase angle difference, calculated as $\alpha'-\alpha$.

LISA is planned to be positioned in an orbit around the Sun, trailing approximately $20\degree$ behind the Earth and maintaining a fixed angle of $60\degree$ with the ecliptic plane.  
It will consist of three satellites separated by a distance of $\sim 2\times10^{6}\,\,{\rm km}$, which serves as the armlength $L_{\rm LISA}$ of the detector. 
These three satellites will complete one orbit in the detector plane over the course of one year. 
The operating period and detection frequency band of LISA are similar to those of TianQin, which opens up the possibility of establishing a network of TianQin + LISA. 

The triangular formation of the three satellites faces challenges in maintaining an equilateral shape due to detector motion, making it difficult to construct an equal-arm Michelson. 
This poses a challenge in canceling the laser noise. 
An effective solution to address this issue is by utilizing \ac{TDI} combination~\cite{Tinto:1999yr}. 
The primary \ac{TDI} channels, namely $\rm X$, $\rm Y$, and $\rm Z$, can be established through each satellite and its adjacent link. 
It is important to note that the angle between these three channels is $60\degree$, resulting in a correlation. 
To address this correlation and eliminate its impact, we can construct two orthogonal and independent channels based on the three primary \ac{TDI} channels. 
One possible option is to incorporate the $\rm A/E$ channels, which are commonly combined with the totally symmetrized channel $\rm T$~\cite{Prince:2002hp}. 
Unlike the $\rm A/E$ channels, $\rm T$ channel is specifically constructed to be insensitive to \acp{GW} in the low-frequency band, making it useful for monitoring channel noise in \ac{GW} detection.  
The channel group $\rm AET$ can be constructed as follows~\cite{Vallisneri:2007xa}: 
\bea
\label{eq:channel_AET}
\nn
{\rm A}&=&\frac{1}{\sqrt{2}}({\rm Z}-{\rm X}),\\
\nn
{\rm E}&=&\frac{1}{\sqrt{6}}({\rm X}-2{\rm Y}+{\rm Z}),\\
{\rm T}&=&\frac{1}{\sqrt{3}}({\rm X}+{\rm Y}+{\rm Z}).
\eea

\subsection{Detection measurement}
The \ac{SGWB} signal $h_{I}(t)$ in the detector channel $I$ is the convolution of the metric perturbations $h(t,\vec{x})$ and the impulse response $\mathbb{D}^{ab}(t,\vec{x})$~\cite{Romano:2016dpx}. 
Although the channel response will inevitably change due to the detector motion, we can simplify the measurement process without significantly compromising its accuracy by restricting our analysis to the short time scale of $[t_{0}-T/2,t_{0}+T/2]$: 
\bea
\label{eq:ht_sgwb}
\nn
h_{I}(t,t_{0})&=&\mathbb{D}^{P}_{I}[t,\vec{x}(t_{0})]*h_{P}[t,\vec{x}(t_{0})]\\
\nn
&=&\sum_{P=+,\times}\int_{-\infty}^{\infty}{\rm d}f
\int_{S^{2}}{\rm d}\hat{\Omega}_{\hat{k}}
F_{I}^{P}(f,\hat{k},t_{0})\widetilde{h}_{P}(f,\hat{k})\\
&&\times e^{{\rm i}2\pi f[t-\hat{k}\cdot\vec{x}(t_{0})/c]},
\eea
where $\vec{x}$ denotes the position of the measurement at time $t$. 
The response function of channel is the double contraction of the channel tensor and the polarization tensor~\cite{Cornish:2001qi}. 
For more details, interested readers are referred to~\cite{Liang:2022ufy}. 
In this paper, we will not extensively discuss these details but instead proceed to present the response function of the $\rm AET$ channel group:
\bw
\bea
\label{eq:F_aet}
\nn
F_{\rm A}^{P}(f,\hat{k},t_{0})&=&\frac{1}{\sqrt{2}}\big[F_{\rm Z}^{P}(f,\hat{k},t_{0})
e^{-{\rm i}2\pi f\hat{k}\cdot\overrightarrow{A_{0}C_{0}}(t_{0})/c}-F_{\rm X}^{P}(f,\hat{k},t_{0})\big],\\
\nn
F_{\rm E}^{P}(f,\hat{k},t_{0})&=&\frac{1}{\sqrt{6}}
\big[F_{\rm X}^{P}(f,\hat{k},t_{0})-2F_{\rm Y}^{P}(f,\hat{k},t_{0})
e^{-{\rm i}2\pi f\hat{k}\cdot\overrightarrow{A_{0}B_{0}}(t_{0})/c}
+F_{\rm Z}^{P}(f,\hat{k},t_{0})e^{-{\rm i}2\pi f\hat{k}\cdot\overrightarrow{A_{0}C_{0}}(t_{0})/c}\big],\\
F_{\rm T}^{P}(f,\hat{k},t_{0})&=&\frac{1}{\sqrt{3}}
\big[F_{\rm X}^{P}(f,\hat{k},t_{0})+F_{\rm Y}^{P}(f,\hat{k},t_{0})
e^{-{\rm i}2\pi f\hat{k}\cdot\overrightarrow{A_{0}B_{0}}(t_{0})/c}
+F_{\rm Z}^{P}(f,\hat{k},t_{0})
e^{-{\rm i}2\pi f\hat{k}\cdot\overrightarrow{A_{0}C_{0}}(t_{0})/c}\big],
\eea
\ew
where $A_{0}$, $B_{0}$, and $C_{0}$ denote the satellites of space-borne detector, i.e., laser interference sites in the $\rm X$, $\rm Y$, and $\rm Z$ channels. 
To accurately define the separation vector between two channels or detectors in our subsequent analysis, we establish the vertex $A_{0}$ of the $\rm X$ channel as the reference position for the detector.

In the frequency domain, the \ac{SGWB} signal 
\bea
\label{eq:hf_sgwb}
\nn
\widetilde{h}_{I}(f,t_{0})
&=&\sum_{P=+,\times}\int_{S^{2}}\,{\rm{d}}\hat{\Omega}_{\hat{k}}
F_{I}^{P}(f,\hat{k},t_{0})\widetilde{h}_{P}(f,\hat{k}) \\
&&\times e^{-{\rm i}2\pi f\hat{k}\cdot\vec{x}(t_{0})/c},
\eea
and the \ac{PSD} of the \ac{SGWB} signal can be defined as
\be
\label{eq:hIhJ}
\langle\widetilde{h}_{I}(f,t_{0})\widetilde{h}_{J}^{*}(f',t_{0})\rangle
=\frac{1}{2}\delta(f-f')P_{{\rm h}_{IJ}}(|f|,t_{0}).
\ee
Here, $I=J$ refers to the auto \ac{PSD} of one channel, while $I\neq J$ refers to the cross \ac{PSD} of two channels\footnote{To simplify, $P_{{\rm h}_{II}}$ reduces to $P_{{\rm h}_{I}}$.}. 
Combined with~\eq{eq:hf_sgwb} and~\eq{eq:hIhJ}, we can establish the following connection between $P_{{\rm h}_{IJ}}$ and $\mathscr{P}_{\rm h}$:
\bea
\label{eq:Ph_IJ}
P_{{\rm h}_{IJ}}(f,t_{0})&=&
\int_{S^{2}}\,{\rm{d}}\hat{\Omega}_{\hat{k}}
\mathcal{Y}_{IJ}(f,\hat{k},t_{0})\mathscr{P}_{\rm h}(f,\hat{k}), 
\eea
where the antenna pattern involves both the channel response and the separation vector $\Delta \vec{x}=\vec{x}_{I}-\vec{x}_{J}$ between detectors:
\bea
\label{eq:Y_IJ_def}
\nn
\mathcal{Y}_{IJ}(f,\hat{k},t_{0})&=&\frac{1}{2}\sum_{P=+,\times}
F^{P}_{I}(f,\hat{k},t_{0})F^{P*}_{J}(f,\hat{k},t_{0})\\
&\quad&\times
e^{-{\rm i}2\pi f\hat{k}\cdot[\vec{x}_{I}(t_{0})-\vec{x}_{J}(t_{0})]/c}.
\eea
To describe the directional dependence of the antenna pattern, it is necessary to further expand it in terms of spherical harmonics:
\bea
\label{eq:Y_IJ}
\mathcal{Y}_{IJ}(f,\hat{k},t_{0})=\sum_{l=0}^{\infty}\sum_{m=-l}^{l}a_{IJ}^{lm}(f,t_{0})Y_{lm}^{\ast}(\hat{k}),
\eea
with the multipole coefficient
\bea
a_{IJ}^{lm}(f,t_{0})=\int_{S^{2}}{\rm d}\hat{\Omega}_{\hat{k}}
\mathcal{Y}_{IJ}(f,\hat{k},t_{0}) Y_{lm}(\hat{k}).
\eea
Substituting \eq{eq:P_h} and \eq{eq:Y_IJ} to \eq{eq:Ph_IJ}, the \ac{PSD} of the \ac{SGWB} signal is calculated as the sum of the product of the multipole coefficient and the corresponding multipole moment~\cite{Kudoh:2004he}:
\be
\label{eq:Ph2ap}
P_{{\rm h}_{IJ}}(f,t_{0})=
\sum_{l=0}^{\infty}\sum_{m=-l}^{l}a_{IJ}^{lm}(f,t_{0})p_{lm}(f),
\ee
which allows for the determination of the angular \ac{PSD} by taking into account the contributions from various multipole components. 

\subsection{Signal-to-noise ratio}
In addition to \ac{SGWB} signal $h_{I}$, the output from channel $s_{I}$ is also contributed by the channel noise, denoted as $n_{I}$. 
Hence, the complete expression for the output is given by $s_{I} = h_{I} + n_{I}$. 
The \ac{PSD} of the output can be obtained by summing the signal \ac{PSD} $P_{\rm h}$ and the noise \ac{PSD} $P_{\rm n}$:
\bea
\label{eq:S_IJ}
&&\langle \widetilde{s}_{I}(f,t_{0})\widetilde{s}^{\ast}_{J}(f',t_{0})\rangle
\nn \\
&=&\langle \widetilde{h}_{I}(f,t_{0})\widetilde{h}^{\ast}_{J}(f',t_{0})\rangle
+\langle \widetilde{n}_{I}(f)\widetilde{n}^{\ast}_{J}(f')\rangle
\nn
\\
&=&\frac{1}{2}\delta(f-f')
\big[P_{{\rm h}_{IJ}}(|f|,t_{0})+P_{{\rm n}_{IJ}}(|f|)\big],
\eea
where the channel noise is assumed to be stationary. 
Furthermore, the \ac{SGWB} signal and the channel noise are considered to be uncorrelated.

In the presence of noise, extracting the \ac{SGWB} signal from the output becomes a challenging task. 
It requires careful management and monitoring of the channel noise. 
One effective approach to mitigate the impact of channel noise is to employ cross-correlation between two channels with uncorrelated channel noise. 
Ideally, the cross-correlation should result in a zero-value noise  \ac{PSD} $P_{{\rm n}_{IJ}}$~\cite{Christensen:1992wi,Flanagan:1993ix}. 
Alternatively, for a single space-borne detector, the null channel could be utilized to monitor the channel noise~\cite{Tinto:2001ii,Hogan:2001jn}. 
By utilizing the above methods, the optimal \ac{SNR} for \ac{SGWB} detection~\cite{Liang:2022ufy}
\be
\label{eq:snr_IJ}
\rho_{IJ}(t_{0})=
\sqrt{\frac{2\,T}{1+\delta_{IJ}}
\int_{f_{\rm min}}^{f_{\rm max}}{\rm d}f\,
\frac{|P_{{\rm h}_{IJ}}(f,t_{0})|^{2}}{P_{{\rm n}_{I}}(f)
P_{{\rm n}_{J}}(f)W_{IJ}(f,t_{0})}},
\ee
where $T$ and $[f_{\rm min},f_{\rm max}]$ denote the correlation time and the detection band, respectively. 
In the context of the channel group $\rm AET$ of space-borne detectors, the noise \ac{PSD} is given by
\bw
\bea
\label{eq:Pn_all}
\nn
P_{\rm n_{\rm A/E}}(f)
&=&\frac{2\sin^{2}\big[\frac{f}{f_{\ast}}\big]}{L^{2}}
\bigg[\bigg(\cos\big[\frac{f}{f_{\ast}}\big]+2\bigg)S_{\rm p}(f)+2\bigg(\cos\big[\frac{2f}{f_{\ast}}\big]+2\cos\big[\frac{f}{f_{\ast}}\big]
+3\bigg)\frac{S_{\rm a}(f)}{(2\pi f)^{4}}\bigg],\\
P_{\rm n_{T}}(f)
&=&\frac{8\sin^{2}\big[\frac{f}{f_{\ast}}\big]\sin^{2}\big[\frac{f}{2f_{\ast}}\big]}{L^{2}}
\bigg(S_{\rm p}(f)+4\sin^{2}\big[\frac{f}{2f_{\ast}}\big]\frac{S_{\rm a}(f)}{(2\pi f)^{4}}\bigg),
\eea
\ew
which takes into account detector armlength $L$, optical-path noise $S_{\rm p}$ and acceleration noise $S_{\rm a}$. 
For more detailed information on the parameters of TianQin and LISA, readers can refer to Ref.~\cite{TianQin:2020hid} and Ref.~\cite{Babak:2021mhe}, respectively.
The correction function $W_{IJ}$ plays an important role when the \ac{SGWB} signal exceeds the channel noise~\cite{Liang:2022ufy}.  

If one specifically considers the contribution of different multipole moments to the \ac{SNR}, where each $m$ is derived from the same Gaussian statistics~\cite{LISACosmologyWorkingGroup:2022kbp}, i.e., 
\be
\mathscr{A}_{l}(f)=p_{lm}(f)p^{*}_{lm}(f),
\ee
then based on the definition of \ac{ORF}~\cite{LISACosmologyWorkingGroup:2022kbp}:
\be
\label{eq:Gamma_IJ_lm}
\Gamma_{IJ}^{lm}(f,t_0)
=\frac{a_{IJ}^{lm}(f,t_0)}{\sqrt{4 \pi}}
=\frac{1}{4 \pi}\int_{S^{2}}{\rm d}\hat{\Omega}_{\hat{k}}\,
\mathcal{Y}_{IJ}(f,\hat{k},t_{0}) \frac{Y_{lm}(\hat{k})}{Y_{00}(\hat{k})},
\ee
the $l-$dependent \ac{SNR} can be defined by substituting \eq{eq:Ph2ap} into \eq{eq:snr_IJ}:
\bw
\bea
\label{eq:rho_l_t}
\nn
\rho^{(l)}_{IJ}(t_{0})
&=&\sqrt{\frac{2\,T}{1+\delta_{IJ}}
\int_{f_{\rm min}}^{f_{\rm max}}{\rm d}f\,
\sum_{m=-l}^{l}\frac{a_{IJ}^{lm}(f,t_{0})a_{IJ}^{lm*}(f,t_{0})
p_{lm}(f)p^{*}_{lm}(f)}
{P_{{\rm n}_{I}}(f)P_{{\rm n}_{J}}(f)W^{lm}_{IJ}(f,t_{0})}}\\
&=&
\sqrt{\frac{2\,T}{1+\delta_{IJ}}
\int_{f_{\rm min}}^{f_{\rm max}}{\rm d}f\,
\frac{|\Gamma^{(l)}_{IJ}(f,t_{0})|^{2}
4\pi (2l+1)\mathscr{A}_{l}(f)}{P_{{\rm n}_{I}}(f)
P_{{\rm n}_{J}}(f)}}.
\eea
\ew
Here the $l-$dependent \ac{ORF}
\bea
\label{eq:Gamma_IJ_l}
\Gamma_{IJ}^{(l)}(f,t_{0})
&=&\sqrt{\frac{1}{2l+1}\sum_{m=-l}^{l}\frac{|\Gamma_{IJ}^{lm}(f,t_{0})|^2}{W_{IJ}^{lm}(f,t_{0})}},
\eea
with the correction function 
\bea
\label{eq:W_IJ}
\nn
W_{IJ}^{lm}(f,t_{0})
&=&1+\frac{P^{lm}_{{\rm h}_{I}}(f)P_{{\rm n}_{J}}(f)
+P^{lm}_{{\rm h}_{J}}(f)P_{{\rm n}_{I}}(f)}
{P_{{\rm n}_{I}}(f)P_{{\rm n}_{J}}(f)}\\
\nn
&&+
\frac{P^{lm}_{{\rm h}_{I}}(f)P^{lm}_{{\rm h}_{J}}(f)+
(1-\delta_{IJ})|P^{lm}_{{\rm h}_{IJ}}(f,t_{0})|^{2}}{P_{{\rm n}_{I}}(f)P_{{\rm n}_{J}}(f)}.\\
\eea
The \ac{SNR} formula allows us to analyze and quantify the impact of each multipole component on the \ac{SGWB} detection, taking into account their statistical properties. 

Without loss of generality, the \ac{SNR} corresponding to the total correlation time $T_{\rm tot}$
\be
\label{eq:rho_l}
\rho^{(l)}_{IJ}=
\sqrt{\frac{2\,T_{\rm tot}}{1+\delta_{IJ}}
\int_{f_{\rm min}}^{f_{\rm max}}{\rm d}f\,
\frac{|\bar{\Gamma}^{(l)}_{IJ}(f)|^{2}
4\pi (2l+1)\mathscr{A}_{l}(f)}{P_{{\rm n}_{I}}(f)
P_{{\rm n}_{J}}(f)}},
\ee
where the time-averaged \ac{ORF}~\cite{Liang:2021bde}
\be
\label{eq:Gamma_bar}
\bar{\Gamma}^{(l)}_{IJ}(f)=
\sqrt{\frac{1}{T_{\rm tot}}\int_{0}^{T_{\rm tot}}{\rm d}t_{0}\,
|\Gamma^{(l)}_{IJ}(f,t_{0})|^{2}}.
\ee
In particular, when $l=0$, \eq{eq:rho_l} and \eq{eq:Gamma_bar} correspond to an isotropic \ac{SGWB}. 

\subsection{Sensitivity curve}
Detection \ac{SNR} is commonly used to assess the detection capability of the detector and detector network for the \ac{SGWB}. 
However, in scenarios where one aims to demonstrate the detection sensitivity, the \ac{PLIS} curve serves as a more suitable indicator~\cite{Thrane:2013oya}. 
This type of sensitivity curve, specifically tailored for power-law \ac{SGWB}, offers a robust method to assess the detection sensitivity to \ac{SGWB}, regardless of the intensity level of the \ac{SGWB}. 

To elaborate on the \ac{PLIS} curve, let us begin with the power-law energy spectrum density:
\be
\label{eq:Omega_pl}
\Omega^{(l)}_{\rm gw}(f)=\Omega^{(l)}_{0}(\epsilon)(f/f_{\rm ref})^{\epsilon}|_{\epsilon=\epsilon_{0}},
\ee
where 
$\Omega_{0}$ is normalized by the reference frequency $f_{\rm ref}$, $\epsilon$ denotes the spectral index, and to be consistent with \eq{eq:Omega2Sh} at $l=0$, we have
\be
\label{eq:Omega_l2Sh_l}
\Omega^{(l)}_{\rm gw}(f)
=\frac{2\pi^{2}}{3H_{0}^{2}}f^{3}\sqrt{4\pi (2l+1)\mathscr{A}_{l}(f)}.
\ee
Here, the angular power spectrum $\mathscr{A}_{l}$ of the \ac{SGWB} at different scales $l$ can be calculated using $\Omega^{(l)}_{\rm gw}$, which is, in particular, numerically equal to the energy spectral density $\Omega_{\rm gw}$ of the \ac{SGWB} when $l = 0$. 

Substituting \eq{eq:Omega_l2Sh_l} into \eq{eq:rho_l}, and setting the \ac{SNR} equal to the threshold $\rho_{0}$ required for detection, one can determine the energy spectrum density $\Omega^{(l)}_{\rm gw}$ for each index $\epsilon$ through inverse solving. 
At each frequency, the maximum value of $\Omega^{(l)}_{\rm gw}$ corresponding to a specific index $\epsilon$ can be selected, enabling the construction of the \ac{PLIS} curve:
\be
\label{eq:Omega_PLI}
\Omega_{\rm PLIS}(f)={\rm max}_{\epsilon}[\Omega^{(l)}_{\rm gw}(f)].
\ee
In particular, if we assume that the \ac{SGWB} is significantly weaker than the channel noise, the process of inverse solving $\Omega^{(l)}_{\rm gw}$ can be simplified by extracting $\Omega^{(l)}_{0}$ from the integral:
\be
\Omega^{(l)}_{0}=\rho_{0}
\left[\frac{2\,T_{\rm tot}}{1+\delta_{IJ}}
\int_{f_{\rm min}}^{f_{\rm max}}{\rm d}f
\frac{(f/f_{\rm ref})^{2\epsilon}}{[\Omega_{{\rm n}_{IJ}}^{(l)}(f)]^{2}}\right]^{-1/2},
\ee
where the $l-$dependent effective sensitivity
\be
\label{eq:Omega_n}
\Omega^{(l)}_{{\rm n}_{IJ}}(f)=
\frac{2\pi^{2}}{3H_{0}^{2}}f^{3}\frac{\sqrt{P_{{\rm n}_{I}}(f)P_{{\rm n}_{J}}(f)}}{\bar{\Gamma}^{(l)}_{IJ}(f)}.
\ee
If the energy spectrum density $\Omega^{(l)}_{\rm gw}$ of a multipole moment is located above the \ac{PLIS} curve, it indicates that the corresponding \ac{SNR} is expected to exceed the preset threshold. 
In such cases, the multipole moment is likely to be detected successfully. 
Conversely, if $\Omega^{(l)}_{\rm gw}$ falls below the \ac{PLIS} curve, it suggests that the detection is highly likely to fail. 
Therefore, the lowest point on the \ac{PLIS} curve denotes the minimum detectable energy spectrum density, which is referred to as ``sensitivity'' when discussing the detection capability. 
Furthermore, the total \ac{SNR} with multiple pairs of channels, denoted as $\rho_{\rm tot}$, is the root sum square of each individual \ac{SNR}: $\rho_{\rm tot}=\sqrt{\sum_{IJ}\rho_{IJ}^{2}}$~\cite{Liang:2021bde}. Then the total effective sensitivity for single detectors or detector networks is given by
\bea
\label{eq:Omega_n_tot}
\nn
\Omega^{(l)}_{{\rm n}_{\rm tot}}(f)&=&
\left\{\sum_{IJ}\left[\frac{1}{\Omega^{(l)}_{{\rm n}_{IJ}}(f)}\right]^{2}\right\}^{-1/2}\\
&=&
\frac{2\pi^{2}}{3H_{0}^{2}}f^{3}
\left\{\sum_{I,J}\frac{|\bar{\Gamma}^{(l)}_{IJ}(f)|^{2}}{P_{{\rm n}_{I}}(f)P_{{\rm n}_{J}}(f)}\right\}^{-1/2}.
\eea

\section{Implementation for detection improvement}\label{sec:result_I}
A single space-borne detector is capable of detecting the monopole ($l=0$), quadrupole ($l=2$), and hexadecapole ($l=4$) of the \ac{SGWB}, but encounters challenges in detecting additional multipole moments~\cite{Cornish:2001hg,Kudoh:2004he,LISACosmologyWorkingGroup:2022kbp}. 
Hence, this section aims to investigate whether a detector network can expand the range of detectable multipole moments.

Let us start with the \ac{ORF}, which is conveniently used to characterize the correlation of the \ac{SGWB} between different detector channels~\cite{Christensen:1992wi,Flanagan:1993ix}. 
To maintain the inherent consistency in the analysis, we initially assume that the separation between detectors, denoted as $|\Delta \vec{x}|$, is much larger than the armlength $L$ of the detector. 
This assumption allows us to utilize the approximation $f\ll c/(2\pi|\Delta \vec{x}|)\ll c/(2\pi L)$. 
Furthermore, we focus on channel combinations denoted as $\{\rm AA',AE',EA',EE'\}$. 
Here, the symbol $'$ is employed to distinguish between the two detectors. 
This selection ensures the \ac{ORF}'s invariance with respect to rotations of the detector channels within the detector plane~\cite{Seto:2020mfd}. 
Assuming that the $\rm A/E$ channels of the same detector share the same noise \ac{PSD}, the total \ac{ORF} of $\{\rm AA',AE',EA',EE'\}$ at the low-frequency limit
\bea
\label{eq:ORF_l_lf}
\nn
\Gamma_{\rm tot}^{(l)}(f)\propto\left\{
\begin{array}{lr}
	f^{2},&l=0,2,4 \\
	\\
	\frac{|\Delta \vec{x}|}{c}\,f^{3},&l=1,3,5\\
	\\
	\frac{|\Delta \vec{x}|^{l-4}}{c}\,f^{l-2},& l\geq 6
\end{array},\quad f\ll c/(2\pi|\Delta \vec{x}|).
\right.\\
\eea
From~\eq{eq:ORF_l_lf}, it becomes clear that, aside from the instances where $l=1$, 2, and 4, the presence of detector separation can significantly enhance the \ac{ORF} for other multipole moments at lower frequencies. 
Furthermore, as the order of the multipole moment increases, the improvement facilitated by detector separation becomes increasingly noticeable. 
However, the total \ac{ORF} is suppressed by detector separation at high-frequency limit:
\be
\label{eq:ORF_l_hf}
\Gamma_{\rm tot}^{(l)}\propto \frac{1}{|\Delta \vec{x}|},\quad f\gg c/(2\pi|\Delta \vec{x}|).
\ee
Additional details regarding \eq{eq:ORF_l_lf} and \eq{eq:ORF_l_hf} are provided in Appendix~\ref{appen:ORF}. 
Furthermore, to mitigate the influence of the frequency-dependent factor introduced by the \ac{TDI}, one can normalize the \ac{ORF} by $C_{\rm TDI}(f)=4\sin\left[2\pi fL/c\right]\sin\left[2\pi fL'/c\right]$, where $L=L'$ for a single detector~\cite{Cornish:2001bb,Liang:2022ufy}. 
Unless otherwise specified, we will adopt this presentation method for the \ac{ORF} in this paper.

The space-borne detectors can allow for detector separation with a wide selectable range~\cite{TianQin:2015yph,LISA:2017pwj,Hu:2017mde}, which ensures that the space-borne detector network meets the condition stated in $f\ll c/(2\pi|\Delta \vec{x}|)\ll c/(2\pi L)$. As an illustrative example, let us consider the TianQin + LISA network~\cite{Liang:2021bde}. 
The separation $|\Delta \vec{x}|$ between TianQin and LISA is approximately equal to $2\sin(\kappa/2)\,\,\rm A.U.$~\cite{Liu:2022umx}, where $\kappa$ represents the angle at which LISA is positioned behind the Earth. 
To demonstrate the influence of detector separation on the \ac{ORF}, we select specific values of $\kappa$: $0\degree$, $6\degree$, $20\degree$, and $60\degree$, corresponding to $|\Delta \vec{x}|$ values of $\rm 0\,\,A.U.$, $\rm 0.1\,\,A.U.$, $\rm 0.35\,\,A.U.$, and $\rm 1\,\,A.U.$, respectively. 
For visual clarity, we maintain TianQin and LISA perpendicular to each other, i.e., $\theta_{\rm r}=\pi/2$, with the unit separation vector $\Delta\hat{x}$ $(\theta= 0, \phi = 0)$.

\begin{figure}[htbp]
\centering
\includegraphics[height=6cm]{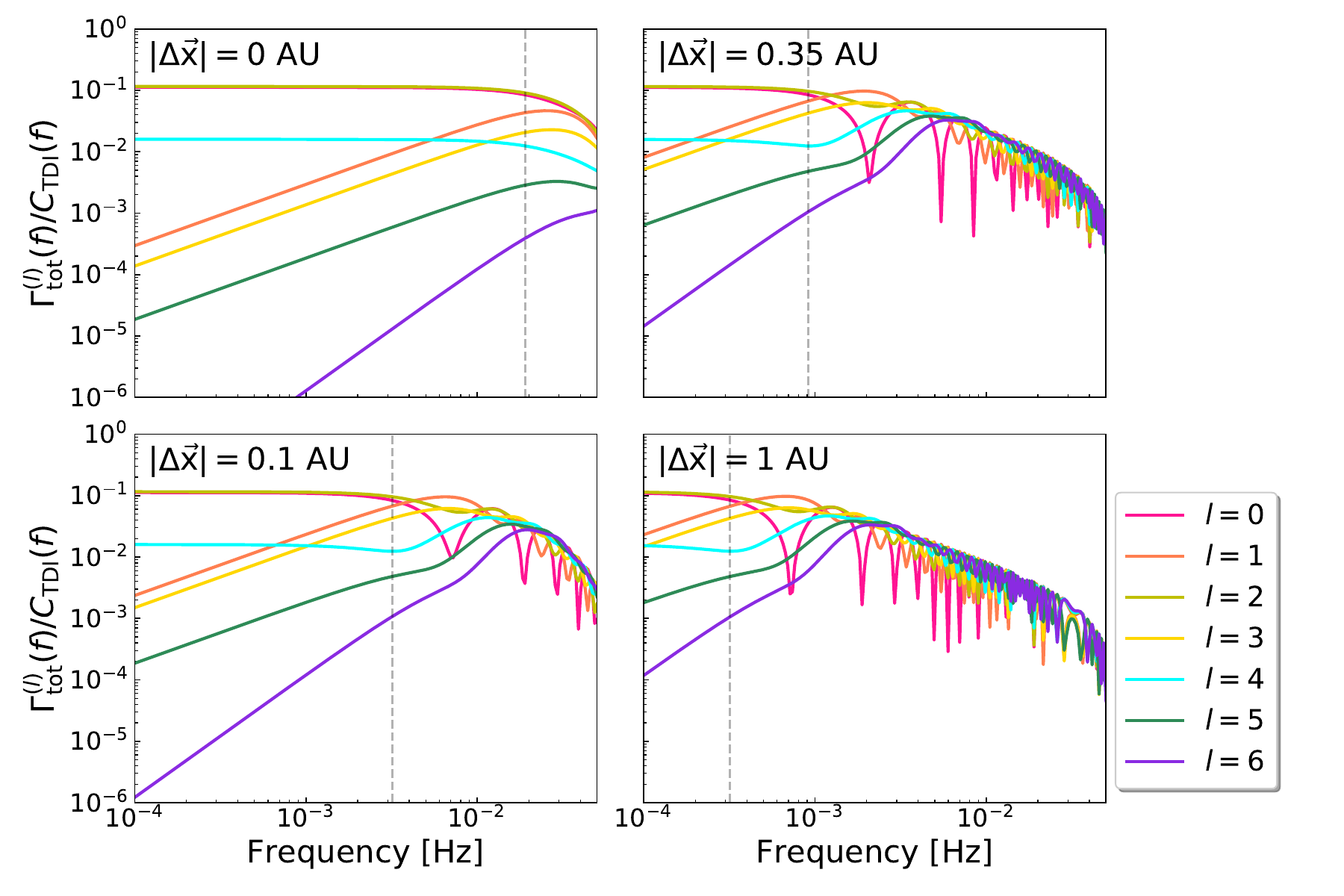}
\caption{The \ac{ORF} of the TianQin + LISA 
network is evaluated for different detector separations: $|\Delta \vec{x}|=0\,\,{\rm A.U.}$, $0.1\,\,{\rm A.U.}$, $0.35\,\,{\rm AU}$, and $1\,\,{\rm A.U.}$. The dashed gray line represents the characteristic frequency, below which the low-frequency limit fails. It is worth noting that when the separation $|\Delta \vec{x}|$ is equal to 0 A.U., the characteristic frequency turns to $c/(2 \pi L)$.}
\label{fig:ORF_tl_l}
\end{figure}

Figure.~\ref{fig:ORF_tl_l} shows the \ac{ORF} of the TianQin + LISA network for varying detector separations, demonstrating a close correspondence with~\eq{eq:ORF_l_lf} and~\eq{eq:ORF_l_hf}. 
For multipole moments with $l=0$, 2 and 4, increasing detector separation has negligible impact on the \ac{ORF} at low frequencies. 
Specially, within the $\rm mHz$ frequency band, the \ac{ORF} for $l=0$ and 2 is suppressed due to increased detector separation. 
On the other hand, for $l=1$, 3, 5 and 6, increasing detector separation can provide a positive effect on the \ac{ORF} at low frequencies.

\begin{figure*}[htbp]
	\centering
	\includegraphics[height=5.8cm]{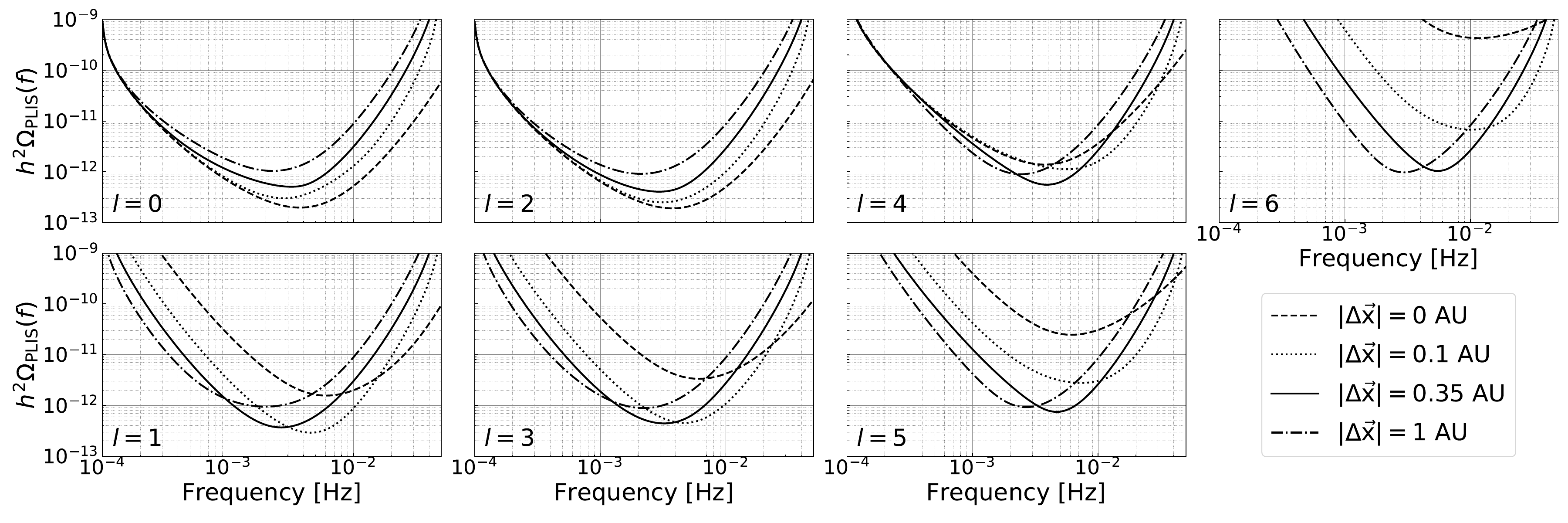}
	\caption{\ac{PLIS} curve of the TianQin + LISA network for each multipole moment, considering different detector separations $|\Delta \vec{x}|$. Note that the Hubble constant $h=0.674$~\cite{Planck:2018vyg}.}
	\label{fig:Omega_tl_l}
\end{figure*}

The next step involves incorporating detector noises to determine the sensitivity. 
In \fig{fig:Omega_tl_l}, we present the \ac{PLIS} curve for different detector separations, considering an \ac{SNR} of $\rho_{0}=1$ and a correlation time $T_{\rm tot}=0.5\,\,{\rm year}$. 
For multipole moments with $l=0$, 2, and 4, it can be concluded that the detector separation does not significantly improve detection sensitivity. 
However, for the remaining cases, introducing detector separation results in enhanced detection sensitivity.  
For multipole moments with $l=1$ and $3$, the minimum detectable energy density can be reduced from $\sim 10^{-12}$ to $\sim 10^{-13}$. 
Additionally, for $l=5$, and 6, this value can be further reduced from $\sim 10^{-11}$ and $\sim 10^{-10}$ respectively to less than $10^{-12}$. 
It is worth noting that, among all the scenarios shown in \fig{fig:Omega_tl_l}, the detector separation of $\rm 0.35\,\,AU$ (corresponding to $\kappa=20\degree$) enables relatively optimal detection sensitivity. 

\section{Realistic case study}\label{sec:result_II}
\begin{table}
	\begin{center}
		\caption{The combination of channels for correlation in a single TianQin, LISA, and the TianQin I+II, TianQin +LISA network. The symbol $^\prime$ is employed to distinguish the channels of different detector in a detector network.}
		\label{tab:channel}
		\setlength{\tabcolsep}{1mm}
		\renewcommand\arraystretch{1.5}
		\begin{tabular}{*{3}{c}}
			\hline
			\hline
			detector/detector network & channel combination                  \\
			\hline
			TianQin or LISA      &  $\rm AA$,$\rm AE$,$\rm AT$,$\rm TT$                       \\
			\multirow{2}*{TianQin I+II or TianQin + LISA} &  $\{\rm AA'$,$\rm AE'$,$\rm EA'$,$\rm EE'$\}    \\
			& $\{\rm AT'$,$\rm ET'$,$\rm TA'$,$\rm TE'\}$,$\{\rm TT'\}$                      \\
			\hline
			\hline  
		\end{tabular}
	\end{center}
\end{table}

In this section, we focus on demonstrating the detection sensitivity to the anisotropic \ac{SGWB} by presenting numerical results for the planned TianQin I+II and TianQin + LISA networks. 
To ensure a comprehensive comparison, we also add the result of single detectors, namely TianQin and LISA individually. 

For a single detector, all possible correlations of the channels can be utilized, including auto-correlation within the same channel and cross-correlation between different channels. 
Since the \ac{ORF} exhibits symmetry in the channels, where $\Gamma_{\rm AA}^{(l)}=\Gamma_{\rm EE}^{(l)}$ and $\Gamma_{\rm AT}^{(l)}=\Gamma_{\rm ET}^{(l)}$, we only need to consider the performance of four pairs of channels. 
When analyzing detector networks and aiming to emphasize the improvement achieved by multiple detectors in comparison to a single detector, the focus shifts to solely considering the cross-correlation between different detectors while disregarding the correlations within each individual detector. 
The channel combinations are classified into three categories based on the presence of null channel: those without null channels (0N), those with one null channel (1N), and those with two null channels (2N). 
These categories are consistently labeled throughout the paper. 
\tab{tab:channel} provides a list of the correlation channel combinations used for our analysis. 
To simplify the notation, we may use shorthand labels such as ``tt" for TianQin I+II, and ``tl" for TianQin + LISA in the following figures and equations. 

\subsection{Overlap reduction function}
\subsubsection{Single detectors}
\begin{figure}[htbp]
	\centering
	\includegraphics[height=6.5cm]{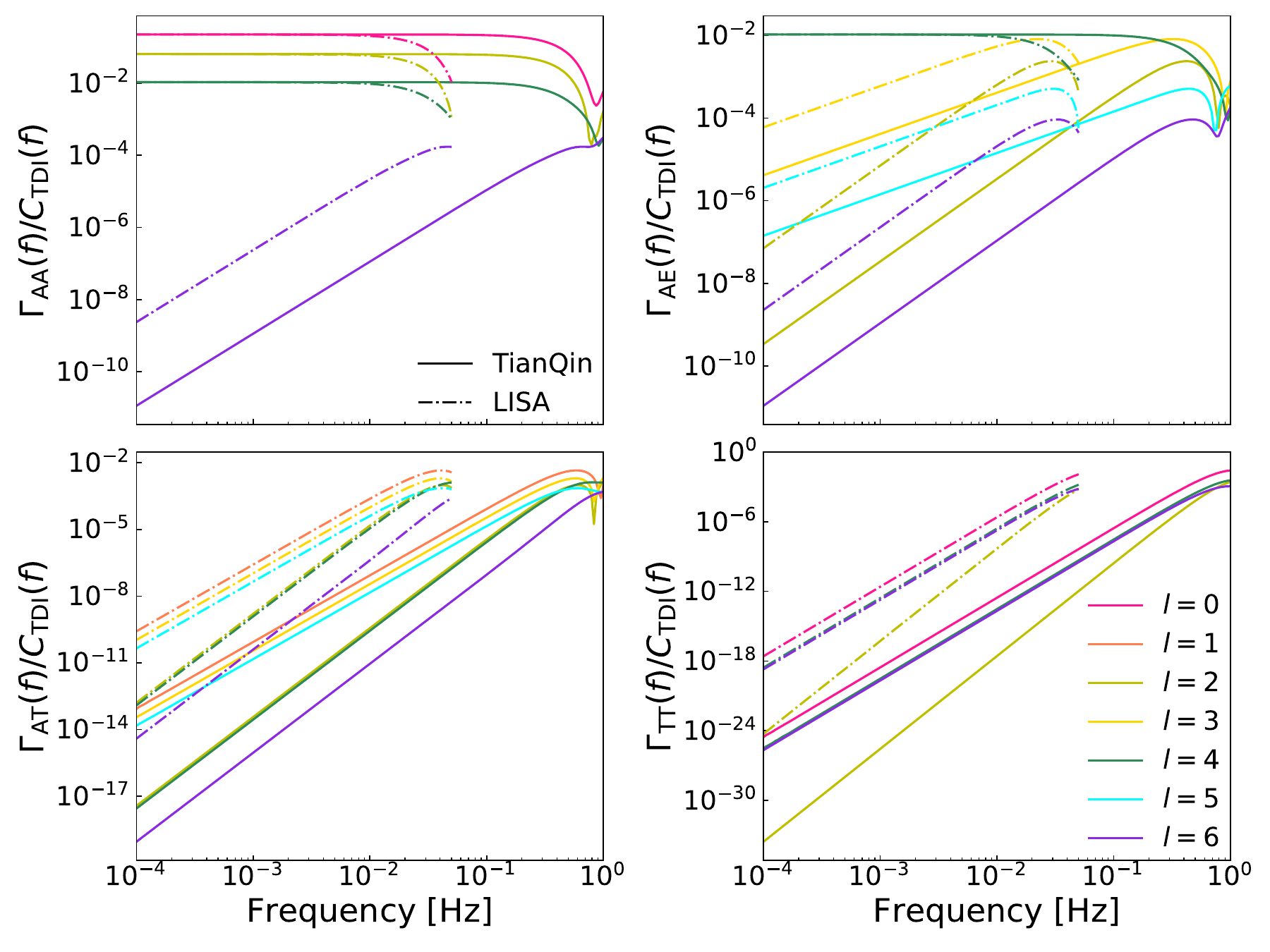}
	\caption{\ac{ORF} for a single TianQin and LISA, taking into account the channel combinations $\rm AA$, $\rm AE$, $\rm AT$, and $\rm TT$. The solid line and the dot-dashed line denote TianQin and LISA, respectively.}
	\label{fig:ORF_lm_TQ}
\end{figure}
Figure.~\ref{fig:ORF_lm_TQ} illustrates the \acp{ORF} of a single TianQin and LISA. 
Due to the symmetry of the spherical harmonic, which states that $Y_{lm}(\hat{k})=(-1)^{l}Y_{lm}(-\hat{k})$, the odd multipole moments in the auto-correlation cancel, denoted as $\Gamma^{(l)}_{II}=(-1)^{l}\,\,\Gamma^{(l)}_{II}$~\cite{LISACosmologyWorkingGroup:2022kbp}. 
Consequently, the auto-correlation of a single channel, represented by $\Gamma_{\rm AA}$ and $\Gamma_{\rm TT}$, is unable to capture odd multipole moments. 
To overcome this limitation, cross-correlation between different channels becomes necessary. 
Even for the dipole ($l=1$), relying solely on the cross-correlation of the $\rm A/E$ channels is insufficient due to the dependency of the antenna pattern $\mathcal{Y}_{IJ}$ on the detector's plane symmetry. 
It is necessary to incorporate the $\rm T$ channel, which serves as a null channel in the \ac{GW} detection~\cite{Seto:2020mfd}. 
By combining the null channel and the signal-sensitive channel, the presence of the dipole can be observed through the cross-correlation between these two channels, represented by $\Gamma_{\rm AT}$. 
Additionally, it is important to consider the limitations of the low-frequency limit for a single detector. 
This approximation fails when the frequency scale reaches the characteristic frequency $f_{\ast}=c/(2\pi L)$, where $L$ represents the armlength of the detector~\cite{Cornish:2001bb}. 
As LISA has a longer armlength compared to TianQin, the \ac{ORF} of LISA will start to decline at lower frequencies than that of TianQin. 

\subsubsection{TianQin I+II network}
\begin{figure}[htbp]
	\centering
	\includegraphics[height=6cm]{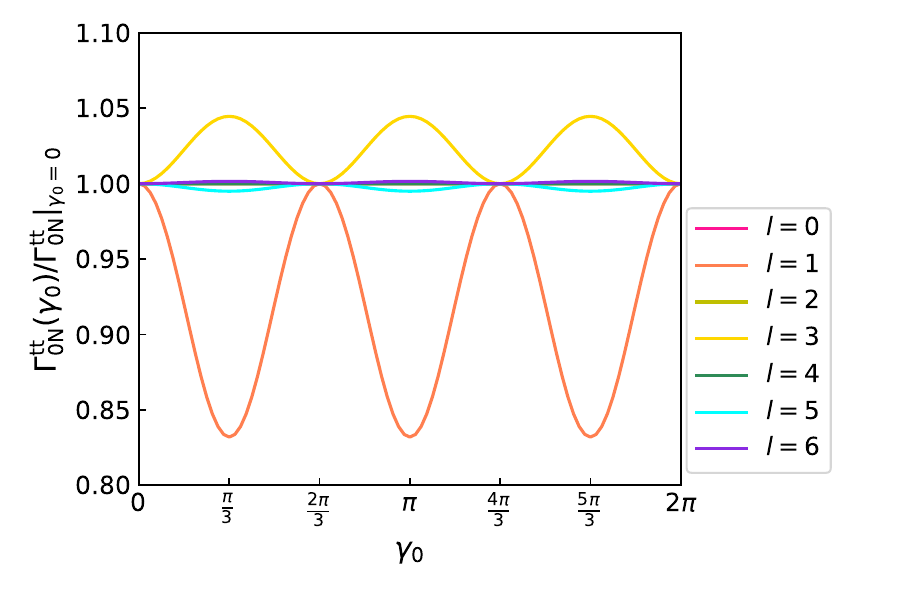}
	\caption{\ac{ORF} of the TianQin I+II network within different initial phase angle difference $\gamma_{0}$ normalized by the case $\gamma_{0}=0$. To comply with the low-frequency limit, we set the frequency $f$ to $0.01\,\,\rm Hz$. It is important to note that for $l=0,2$, and 4, the \acp{ORF} are not affected by $\gamma_{0}$. As a result, they coincide and appear as a single straight line in the plot.}
	\label{fig:ORF_lm_TT_b}
\end{figure}

\begin{figure*}[htbp]
	\centering
	\includegraphics[height=6cm]{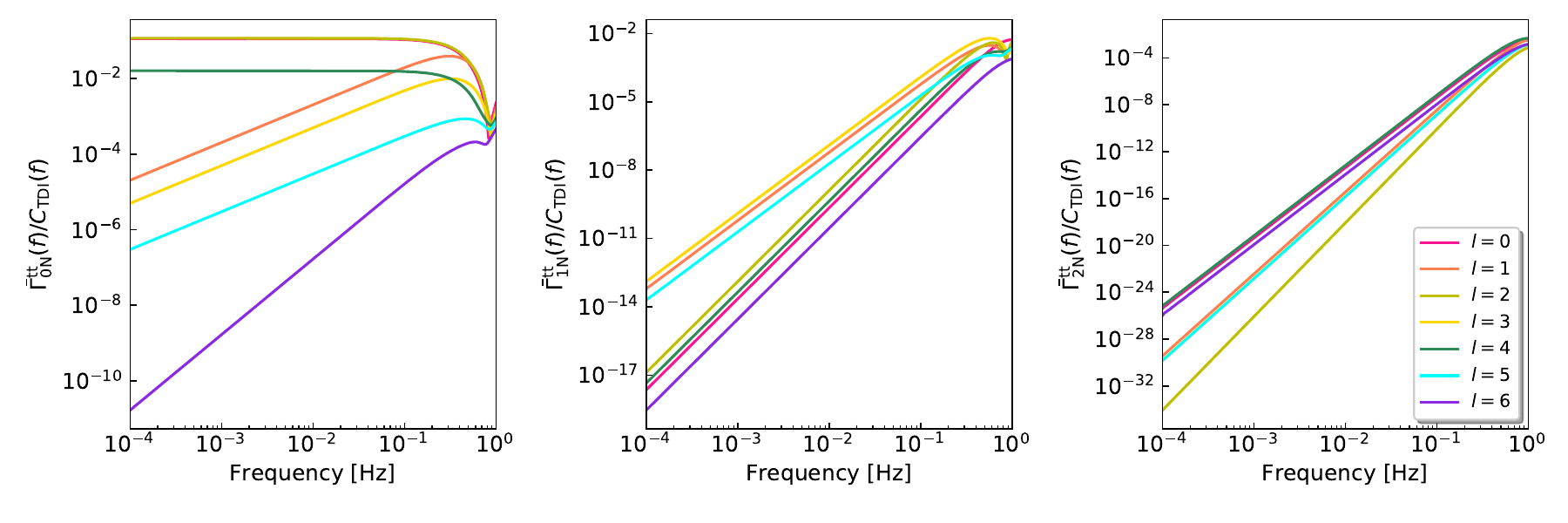}
	\caption{Time-average \ac{ORF} for the TianQin I+II network, where ``$\rm 0N$", ``$\rm 1N$", and ``$\rm 2N$" denote the channel combinations $\{\rm AA'$,$\rm AE'$,$\rm EA'$,$\rm EE'$\}, $\{\rm AT'$,$\rm ET'$,$\rm TA'$,$\rm TE'\}$, and $\{\rm TT'\}$, respectively. }
	\label{fig:ORF_lm_TT}
\end{figure*}

In order to streamline the detection analysis for the TianQin I+II network, the calculation of the \ac{ORF} can be simplified as follows:
\be
\left\{
\begin{array}{lr}
	\Gamma_{\rm 0N}(f)=
	\sqrt{\sum_{IJ}|\Gamma_{IJ}(f)|^{2}},IJ=\{\rm AA',AE',EA',EE'\}\\
	\\
	\Gamma_{\rm 1N}(f)=
	\sqrt{\sum_{IJ}|\Gamma_{IJ}(f)|^{2}},IJ=\{\rm AT',ET',TA',TE'\}\\
	\\
	\Gamma_{\rm 2N}(f)=\Gamma_{\rm TT'}(f)
\end{array},
\right.\\
\ee
where we assume that the A/E/T channels share identical armlengths and secondary noises.

Given the inherent complexity of detector networks compared to single detectors, it is essential to examine the influence of network design on the \ac{ORF}. 
As mentioned in Ref.~\cite{Liang:2021bde}, the difference in initial angles between TianQin and TianQin II, denoted as $\gamma_{0}=\alpha-\alpha'$ (as illustrated in~\fig{fig:Configuration}), can affect the \ac{ORF} of the TianQin I+II network. 
However, for an isotropic \ac{SGWB}, this impact is generally considered negligible under the low-frequency limit~\cite{Seto:2020mfd}. 
In this study, we further investigate the situation when dealing with the anisotropic \ac{SGWB}. 
To narrow down our analysis, we take advantage of the fact that the null channel $\rm T$ is insensitive to \ac{SGWB} at frequencies much lower than the characteristic frequency, which allows us to focus on the $\Gamma_{\rm 0N}$. 

In \fig{fig:ORF_lm_TT_b}, we present the normalized $\Gamma_{\rm 0N}$ for different multipoles with $\gamma_{0}$ set to 0. 
To maintain the validity of the low-frequency limit, the frequency is set to $\rm 0.01\,\,Hz$. 
As explained in Refs.~\cite{Seto:2020mfd,Liang:2021bde}, by summing over four pairs of \ac{TDI} channels, the total \ac{ORF} for the monopole ($l=0$) remains independent of $\gamma_{0}$ in the low-frequency limit. 
This independence primarily stems from the fact that detector separation does not influence the \ac{ORF} for $l=0$, as indicated by~\eq{eq:ORF_l_lf}. 
Conversely, for multipoles with $l$ values other than 0, 2, or 4, the detector separation has a significant impact on the \ac{ORF}. For the TianQin I+II network, there is a rotational symmetry with a period of $2\pi/3$ concerning the phase angle $\gamma_{0}$. 
Consequently, the detector separation undergoes a periodic change with a period of $2\pi/3$. 
\fig{fig:ORF_lm_TT_b} presents the variations within this period in the \ac{ORF} of the TianQin I+II network. 
For $l=0$, 2, and 4, the \ac{ORF} remains unchanged regardless of the value of $\gamma_{0}$. 
However, for $l=1$, 3, 5, and 6, the \ac{ORF} exhibits periodic fluctuations with a period of $2\pi/3$. 
It can be observed that the influence of the phase angle on the \ac{ORF} decreases as the order increases. 
Even for the dipole, the range of \ac{ORF} only exhibits a modest difference of approximately 20\% between the maximum and minimum values. 
To maximize the \ac{ORF} of the dipole, we set the initial phase angle difference $\gamma_{0}$ to 0 and illustrate the time-average \ac{ORF} of the TianQin I+II network in \fig{fig:ORF_lm_TT}. 
Due to the non-coplanarity of TianQin and TianQin II, the \ac{ORF} of the TianQin I+II network for dipoles can be considerably better than that of a single TianQin. However, the separation between TianQin and TianQin II can be even smaller than their armlengths. Therefore, the improvement in \ac{ORF} resulting from the separation will be subject to certain limitations.

\begin{figure*}[htbp]
	\centering
	\includegraphics[height=6cm]{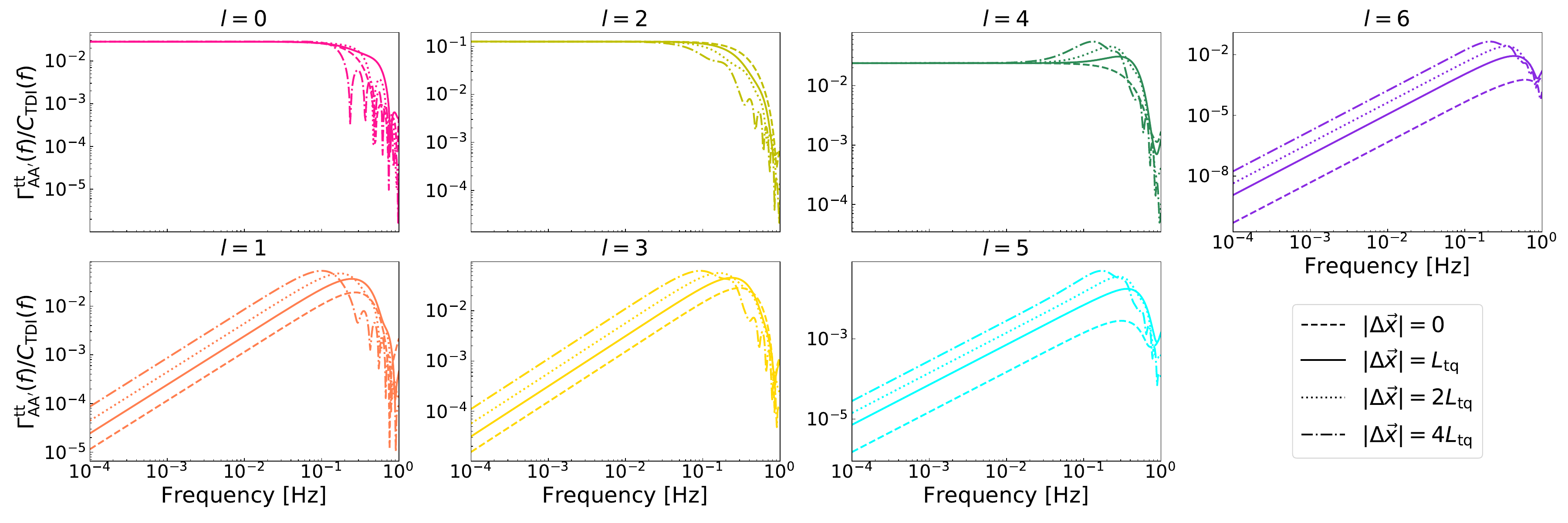}
	\caption{Effect of the detector separation $|\Delta \vec{x}|$ on \ac{ORF} for the TianQin I+II network, where we conducted an analysis with $\alpha=\alpha'=0$.}
	\label{fig:ORF_TT_dx}
\end{figure*}

In order to demonstrate the impact of separation on the \ac{ORF} of the TianQin I+II network, we performed simulations with different separations between TianQin and TianQin II while keeping them perpendicular to each other. \fig{fig:ORF_TT_dx} depicts the corresponding \ac{ORF} for the $\rm A$ channel of TianQin and the $\rm A'$ channel of TianQin II. 
As the separation increases, the characteristic frequency experiences continuous suppression. 
In frequency bands well below the characteristic frequency, the increase in separation does not have a noticeable effect on the \ac{ORF} for $l=0$, 2, and 4. 
However, for the remaining multipole moments, particularly when $l=6$, increasing the separation leads to a significant improvement in the \ac{ORF}. 

\subsubsection{TianQin + LISA network}
\begin{figure}[htbp]
    \centering
    \includegraphics[height=8.8cm]{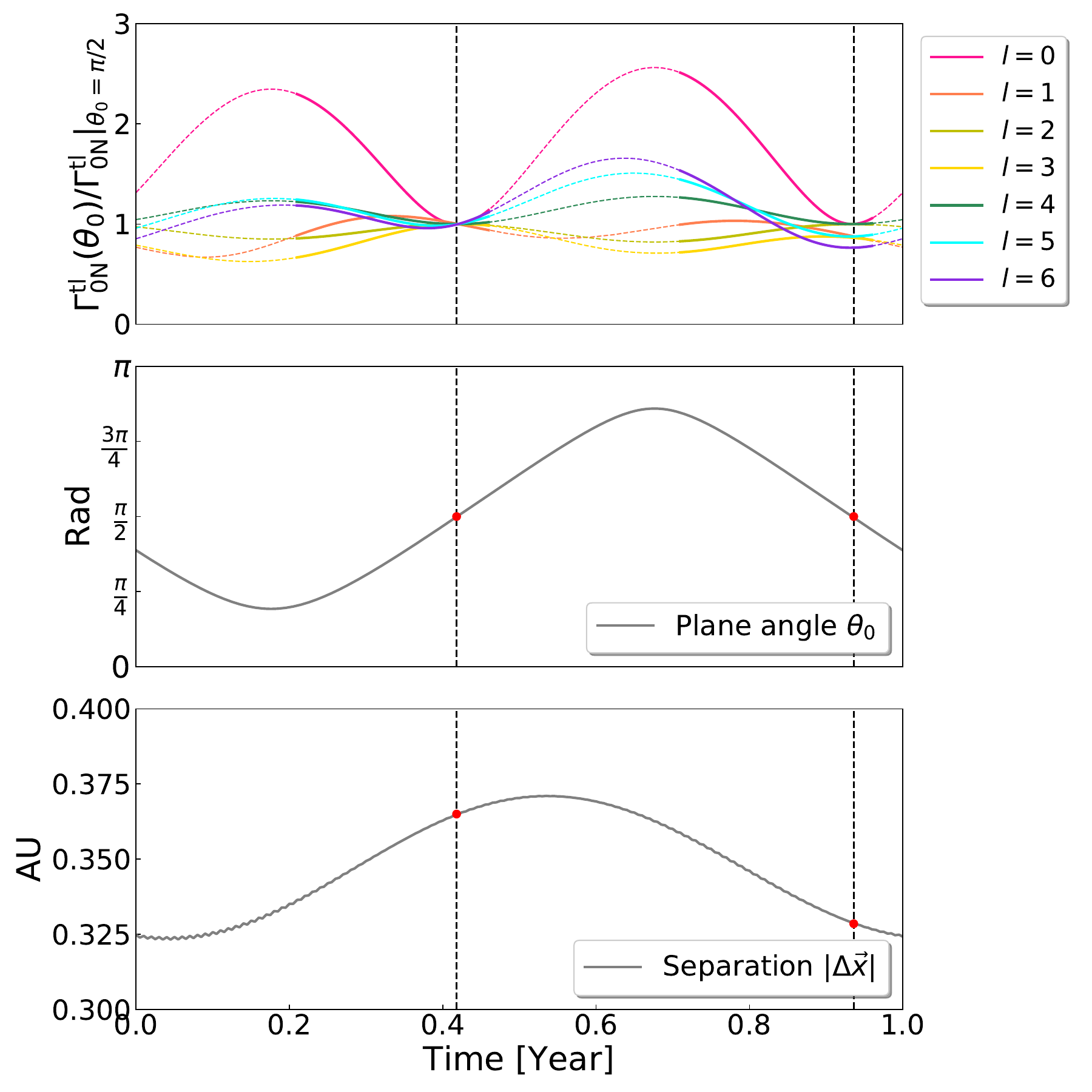}
    \caption{The \ac{ORF} of the TianQin + LISA network undergoes time-dependent changes, which is normalized by comparing them to the case when the two detector planes are perpendicular for the first time. The dashed lines in the plot correspond to two specific time points when the two detector planes are perpendicular. To ensure the validity of the analysis, the frequency is set to $1\times10^{-5}\,\,\rm Hz$, which prevents the low-frequency limit from failing. It is crucial to emphasize that in this paper, the detector separation is defined as the distance between the reference positions of the two detectors, specifically the vertices of the X channel. It is not simply determined by a direct connection between the centers of the two detectors. As a result, the detector separation in the TianQin + LISA network will experience changes over time.}
    \label{fig:ORF_lm_TL_t}
\end{figure}
As for the TianQin + LISA network, it is important to consider the difference in orbital periods between TianQin and LISA. 
The orbital period of LISA is approximately 100 times longer than that of TianQin, allowing LISA to effectively remain stationary during one orbital period of TianQin. 
This averaging effect helps mitigate the impact of the initial phase angle difference. 
However, it is important to note that the plane angle between TianQin and LISA will change over time, necessitating closer attention to the effect of the plane angle on the \ac{ORF} for the TianQin + LISA network. 

In~\fig{fig:ORF_lm_TL_t}, we present the normalized \ac{ORF} of the TianQin + LISA network, with respect to the case where TianQin is perpendicular to LISA. 
Due to the working mode of TianQin, TianQin and LISA simultaneously detect \ac{SGWB} for only half of the year, while the dotted line in the graph corresponds to the period when TianQin is off duty. 
As shown in \fig{fig:Configuration}, TianQin is fixed at an angle of $-4.7 \degree$ relative to the ecliptic plane, while LISA maintains $60 \degree$ with respect to the ecliptic as it orbits the Sun. As a result, there is a varying plane angle between the two detectors, ranging from $35 \degree$ to $155 \degree$, with the minimum and maximum values asymmetrically located around $90 \degree$. Due to this non-symmetrical configuration, differences can arise in the \ac{ORF} of the TianQin + LISA network between the two annual peaks. Specifically, considering the case where $l=0$, the \ac{ORF} exhibits lower values when the detector planes are closer to being perpendicular to each other. Consequently, there is a notable different in the magnitudes of the two peaks within the \ac{ORF}. 
Furthermore, it becomes evident that setting the plane angle $\theta_{0}=\pi/2$ (perpendicular design) is not the optimal choice for detecting the monopole ($l=0$) and hexadecapole ($l=4$). 
However, the situation varies for other multipole moments. 
In particular, for the quadrupole ($l=2$), the best results are achieved when the two detectors are positioned perpendicular to each other. 
It is crucial to highlight that the detector separation of the TianQin + LISA network varies over time. 
After six months, when the two detectors become perpendicular again, the detector separation will no longer be the same as it was half a year ago. 
Consequently, the \acp{ORF} at two perpendicular times are not exactly equal, particularly for multipole moments with $l=1$, 3, 5, and 6. 

\begin{figure}[htbp]
	\centering
	\includegraphics[height=6.5cm]{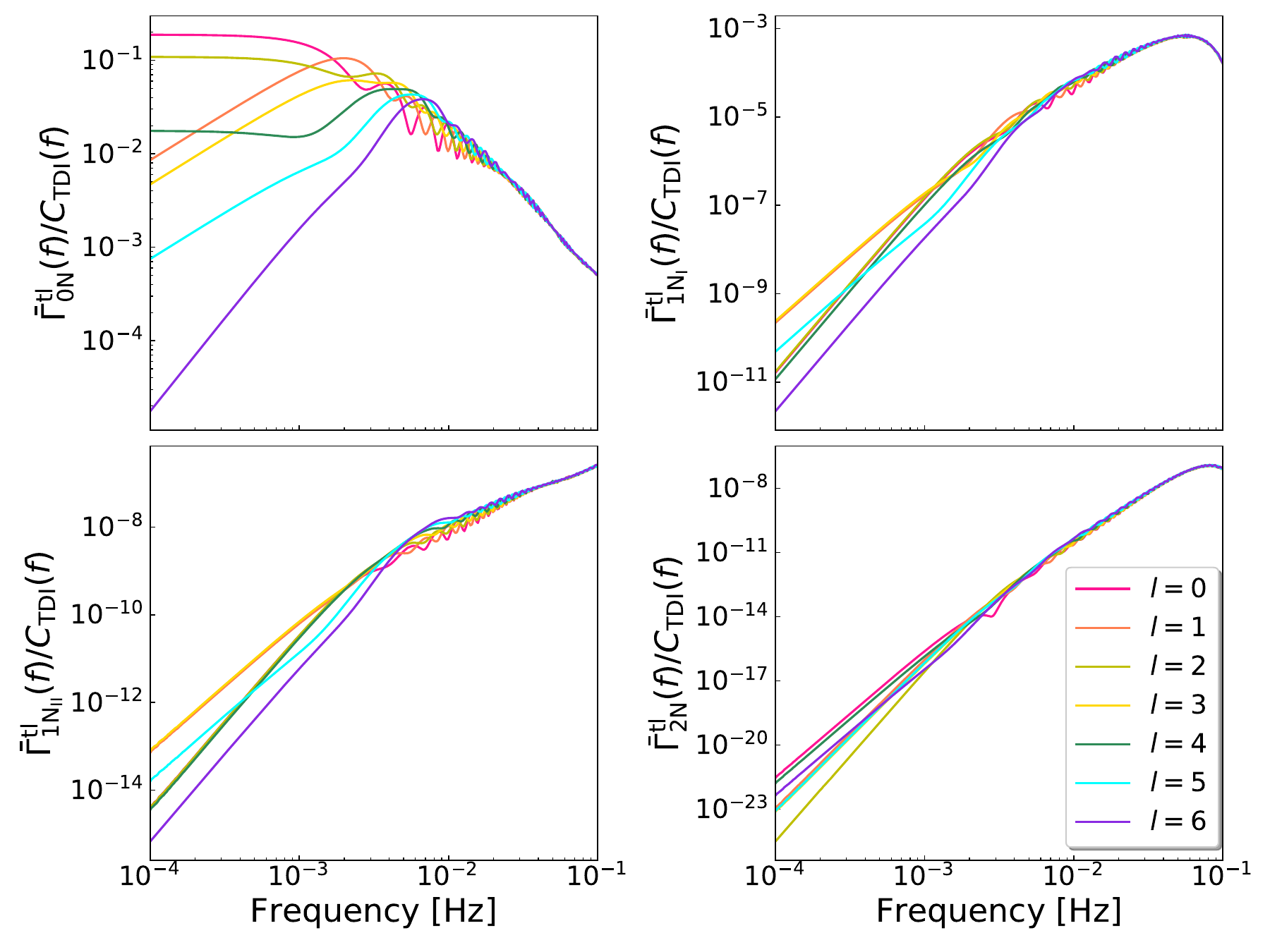}
	\caption{Time-average \ac{ORF} for the TianQin + LISA network, where ``$\rm 0N$", ``$\rm 1N_{I}$", ``$\rm 1N_{II}$", and ``$\rm 2N$" denote the channel combinations $\{\rm AA'$,$\rm AE'$,$\rm EA'$,$\rm EE'$\}, $\{\rm AT'$,$\rm ET'\}$,$\{\rm TA'$,$\rm TE'\}$, and $\{\rm TT'\}$, respectively.}
	\label{fig:ORF_lm_TL}
\end{figure}

Figure.~\ref{fig:ORF_lm_TL} further illustrates the time-average \ac{ORF} of the TianQin + LISA network. 
Due to the differences in armlengths between TianQin and LISA, the channel combination $\{\rm AT'$,$\rm ET'$,$\rm TA'$,$\rm TE'\}$ should be divided into two groups $\{\rm AT'$,$\rm ET'\}$ and $\{\rm TA'$,$\rm TE'\}$, denoted as ``$\rm 1N_{I}$" and ``$\rm 1N_{II}$", respectively. 
Note that, the separation between TianQin and LISA, $|\Delta \vec{x}|\approx 0.35\,\rm A.U.$, is much longer than the armlengths of the two detectors. 
Consequently, the characteristic frequency of the TianQin + LISA network is further lowered to $\sim\rm 1\,\,mHz$~\cite{Liang:2021bde}. 
In the frequency band below $\rm 1\,\,mHz$, noticeable gaps exist between the \acp{ORF} for different multipole moments. 
However, as the frequency increases beyond $\rm 1\,\,mHz$, these gaps tend to significantly narrow. 
When the frequency surpasses $\rm 10\,\,mHz$, the \acp{ORF} for different multipole moments will converge to the similar magnitude, indicating a comparable level of response across the multipole moments. 

\subsection{Antenna pattern}
\begin{figure}[htbp]
	\centering
	\includegraphics[height=11cm]{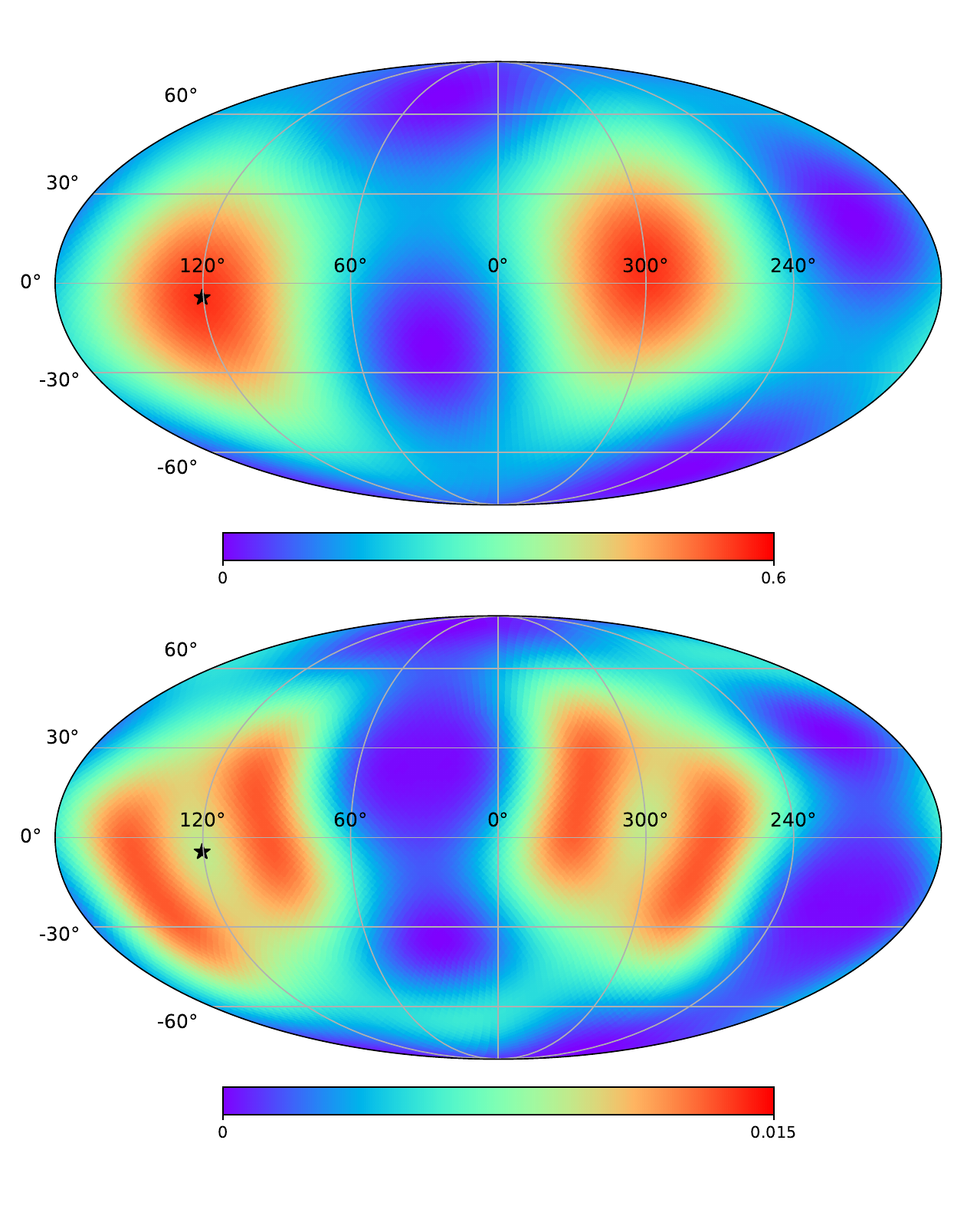}
	\caption{Antenna pattern of TianQin plotted on a Mollweide projection of the sky in ecliptic coordinates. In this representation, the black star signifies the direction of TianQin. Top and bottom panels correspond to frequencies of $10^{-4}\,\,{\rm Hz}$ and $1\,\,{\rm Hz}$, respectively.}
	\label{fig:AP_tq}
\end{figure}

\begin{figure*}[htbp]
	\centering
	\includegraphics[height=11cm]{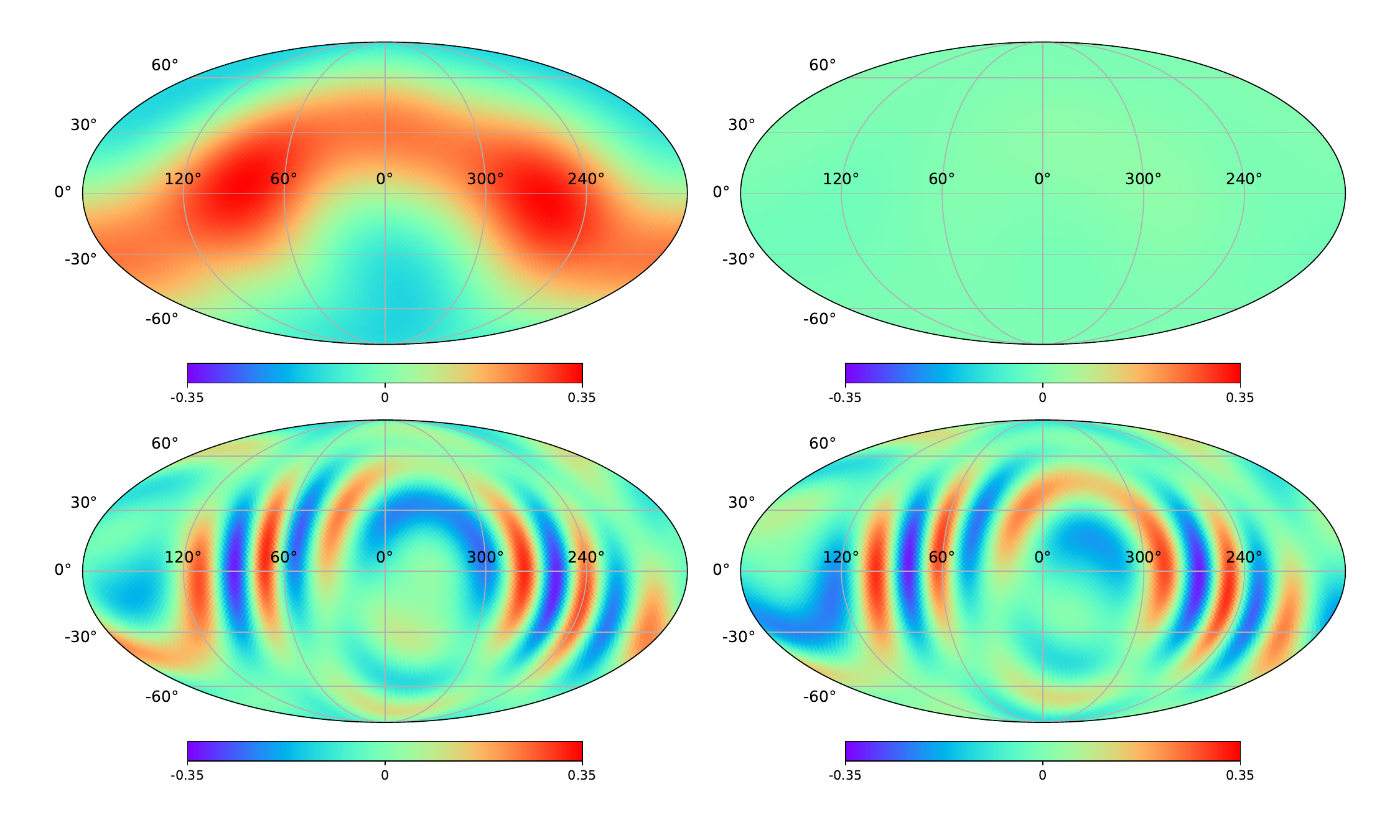}
	\caption{Real and imaginary parts of the antenna pattern of the TianQin + LISA network, which are plotted on a Mollweide projection of the sky in ecliptic coordinates. Top and bottom panels correspond to frequencies of $10^{-4}\,\,{\rm Hz}$ and $10^{-2}\,\,{\rm Hz}$, respectively.}
	\label{fig:AP_tl}
\end{figure*}
The \ac{ORF}, or the cumulative effect of the antenna pattern across the whole sky while considering both positive and negative values, is a crucial aspect. Our objective in this section is to plot the antenna pattern using~\eq{eq:Y_IJ_def}. This visualization can allow us to observe and understand how the channel pair responds to the \ac{SGWB} from different positions.

Due to the $45\degree$ rotation angle difference between the A and E channels of a single detector~\cite{Seto:2020mfd}, it is preferable to focus on the auto-correlation of the A channel within a single detector and the cross-correlation of different A channels in the detector network when visualizing the antenna pattern. To illustrate the antenna pattern, we will use both TianQin and the TianQin + LISA network as examples. For this purpose, we refer to~\fig{fig:ORF_lm_TL_t} and select the reference moment $t=0$. It is important to note that in order to account for the additional factor of $4\,\sin^{2}(f/f_{\ast})$ introduced by \ac{TDI}~\cite{Liang:2022ufy}, the antenna pattern is normalized accordingly.

Figure.~\ref{fig:AP_tq} exhibits the antenna pattern for TianQin. In the low-frequency limit, the hot spots of the antenna pattern align with the fixed direction of TianQin, identified by the coordinates (lon, lat)$=(120.5\degree,-4.7\degree)$ in ecliptic coordinates. Nonetheless, as the frequency exceeds the characteristic frequency of $c/(2\pi L)$, the hot spots shift towards the sides where TianQin is pointing. Similar findings are presented in Ref.~\cite{Romano:2016dpx} regarding LISA. Furthermore, the response function $F^{P}_{I}$ mentioned in~\eq{eq:F_aet} weakens with increasing frequency. This decrease in intensity of the antenna pattern leads to the reduction in the \ac{ORF}.

Unlike the auto-correlation of one channel, the cross-correlation of two channels can introduce the imaginary part of the antenna pattern. In~\fig{fig:AP_tl}, we depict the antenna pattern for the TianQin + LISA network. In the low-frequency limit, the real part of the antenna pattern dominates over the imaginary part, and the \ac{ORF} does not experience significant decline due to the limited positive and negative cancellations within this frequency range. As the frequency surpasses the characteristic frequency of $c/(2\pi |\Delta \vec{x}|)$, the exponential term $e^{-{\rm i}2\pi f\hat{k}\cdot\Delta \vec{x}/c}$ in~\eq{eq:Y_IJ_def} introduces multiple positive and negative oscillations in both the real and imaginary parts of the antenna pattern. This phenomenon is also observed when cross-correlating \ac{LIGO} Hanford and \ac{LIGO} Livingston~\cite{Romano:2016dpx}. Despite no significant change in the peak intensity of the antenna pattern due to the unchanged response function, these oscillations caused by the exponential term lead to a decrease in the \ac{ORF} at this frequency range.

\subsection{Power-law integrated sensitivity curve}
The next step involves demonstrating the detection sensitivity to \ac{SGWB} through the \ac{PLIS} curve. 
To evaluate the sensitivity, we consider all possible channel combinations, including the null channel $\rm T$, which becomes a sensitive channel at high frequencies~\cite{Vallisneri:2007xa}. Furthermore, the operating time for each scenario is set to one year. 
Specifically, for a single TianQin, a single LISA, the TianQin I+II network, and the TianQin + LISA network, the total correlation time is half a year, one year, half a year, four months, and half a year, respectively~\cite{Liang:2021bde}. 

\begin{figure*}[htbp]
	\centering
	\includegraphics[height=7.5cm]{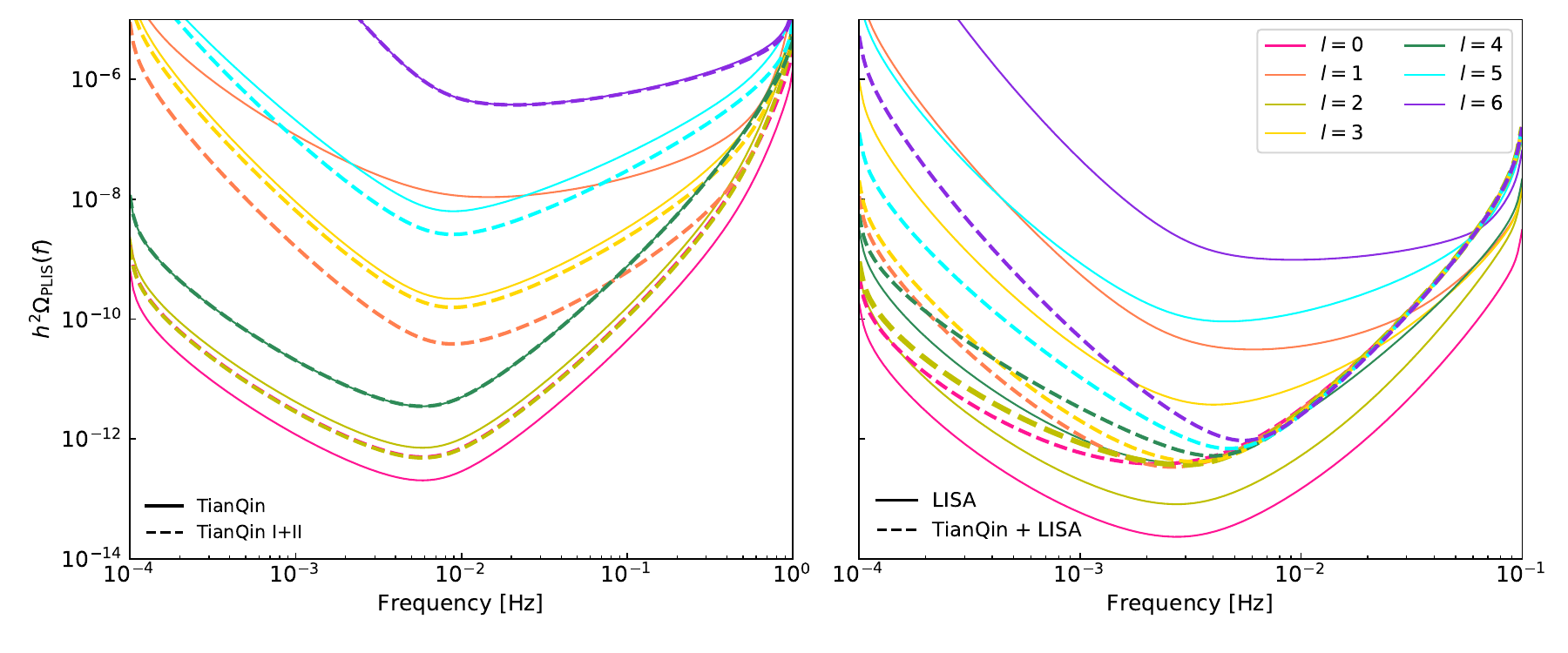}
	\caption{\ac{PLIS} sensitivity to different multipole moments of the \ac{SGWB} for a single TianQin, a single LISA, TianQin I+II, and TianQin + LISA. The \ac{SNR} threshold is set to 1, and the operating time is set to one year. It is important to note that for single detectors, both auto-correlations within the same channel and cross-correlations between different channels are taken into account. However, in the case of a detector network, only the cross-correlation between channels belonging to different detectors is considered.}
	\label{fig:Omega_n_TL}
\end{figure*}

In \fig{fig:Omega_n_TL}, we show the \ac{PLIS} curve for both individual detectors and detector networks using \eq{eq:Omega_PLI}. 
To highlight the improved detection capability of a detector network compared to a single detector in the anisotropic \ac{SGWB} detection, we compare the performance of a single TianQin with the TianQin I+II network, as well as a single LISA with the TianQin + LISA network\footnote{It is crucial to emphasize that the detection of the monopole, which represents the isotropic \ac{SGWB}, cannot benefit from the cross-correlation within a single detector.}. 
The detector network exhibits the comparable detection capability to a single detector for multipole moments with $l=0$, 2, and 4. 
Furthermore, for the remaining multipole moments, combining multiple detectors can significantly improve the detection capability beyond that of a single detector. 
For instance, in the case of detecting the dipole ($l=1$), when TianQin is combined with TianQin II, the detection sensitivity improves from $1\times10^{-8}$ to $4\times10^{-11}$. 
Alternatively, when LISA is combined with TianQin, the detection sensitivity improves from $3\times10^{-11}$ to $3\times10^{-13}$. 
Additional cases are listed in \tab{tab:sensitivity}. 
\begin{table*}
	\begin{center}
		\caption{The sensitivity to different multipole moments for four scenarios: a single TianQin, a single LISA, the TianQin I+II network, and the TianQin + LISA network.}
		\label{tab:sensitivity}
		\setlength{\tabcolsep}{3mm}
		\renewcommand\arraystretch{1.5}
		\begin{tabular}{*{9}{c}}
			\hline
			\hline
			               & 0 & 1 & 2 & 3 & 4& 5& 6 \\
			\hline
			TianQin        & $2\times10^{-13}$ & $1\times10^{-8}$ & $7\times10^{-13}$ & $2\times10^{-10}$ & $4\times10^{-12}$ & $6\times10^{-9}$& $4\times10^{-7}$ \\
			LISA           & $2\times10^{-14}$ & $3\times10^{-11}$ & $8\times10^{-14}$ & $4\times10^{-12}$ & $4\times10^{-13}$ & $9\times10^{-11}$& $1\times10^{-9}$ \\
			TianQin I+II   & $5\times10^{-13}$ & $4\times10^{-11}$ & $5\times10^{-13}$ & $2\times10^{-10}$ & $4\times10^{-12}$ & $3\times10^{-9}$& $4\times10^{-7}$ \\
			TianQin + LISA   & $4\times10^{-13}$ & $3\times10^{-13}$ & $4\times10^{-13}$ & $4\times10^{-13}$ & $5\times10^{-13}$ & $7\times10^{-13}$& $9\times10^{-13}$ \\
			\hline
			\hline
		\end{tabular}
	\end{center}
\end{table*}

\section{Conclusion and discussion}\label{sec:conclusion}
In this paper, we analyze the sensitivity of space-borne detector networks to the anisotropic \ac{SGWB}, and conclude that the separation between detectors plays a significant role in the sensitivity. 
We first derive the dependency of the \ac{ORF} on detector separation in both the low- and the high-frequency limits. 
In the low-frequency limit, the \ac{ORF} of anisotropy exhibits improved performance with increased detector separation, as shown in~\eq{eq:ORF_l_lf}. 
This improvement is particularly notable for high-order multipoles. 
Conversely, in the high-frequency limit, the \ac{ORF} is suppressed as detector separation increases, as indicated in~\eq{eq:ORF_l_hf}. 
Based on these findings, it becomes essential to carefully choose an appropriate detector separation within the network so that the frequency bands with the boosted sensitivity align with the detector's sensitive frequency range. 
For illustration purposes, we employed the TianQin + LISA network with different separations as an example and calculated the \ac{PLIS} curve by combining the \ac{ORF} with the detector noise. 
Our results indicated that compared to scenarios with no separation, 0.1 A.U. separation, and 1 A.U. separation, the current design (0.35 A.U. separation) stands out as the best choice for anisotropy detection.

Furthermore, we expanded our analysis to compare the detection sensitivities of the planned TianQin I+II and TianQin + LISA networks with those of TianQin and LISA alone. 
Our analysis revealed that the detection capabilities of the TianQin I+II network are similar to those of a single TianQin for the monopole ($l=0$), quadrupole ($l=2$), and hexadecapole ($l=4$). 
However, for the remaining multipole moments, the TianQin I+II network surpasses the detection capabilities of a single TianQin by several orders of magnitude. 
Similar conclusions can be drawn when comparing a single LISA with the TianQin + LISA network. 
For example, considering the dipole ($l=1$), which may arise from the motion of the solar system barycenter~\cite{Chung:2022xhv}, the detection sensitivity of the detector network is expected to improve by 2-3 orders of magnitude compared to using a single detector. 
As for the hexacontatetrapole ($l=6$), the detection sensitivity can be enhanced by over 3 orders of magnitude.

Additionally, we explored the impact of the detector plane angle on the performance of the TianQin + LISA network. 
We found that aligning the two detector planes perpendicular to each other is unfavorable for detecting the monopole ($l=0$) and hexadecapole ($l=4$) moments, although this network design does enhance the detection of the quadrupole ($l=2$). 
The exact optimization of the plane angle may vary for different multipole moments, necessitating a case-by-case analysis.

We remark that our analysis is also valid for ground-based detectors that are bound on the Earth. 
However, for a typical separation of thousands of kilometers, the $f_*$ is only a few dozen of hertz, which is not much higher than the lower frequency limit for ground-based detectors. 
Therefore, for a network of LIGO/Virgo or Cosmic Explorer/\ac{ET}, the low-frequency limit fails for most of the sensitive frequency band~\cite{Mentasti:2020yyd,Gupta:2023lga}, and they can not benefit much from larger separations.

Another critical consideration lies in the map-making of the \ac{SGWB}. Thrane et al. utilized spherical harmonics decomposition analysis for extended sources~\cite{Thrane:2009fp}, while another approach involved radiometer analyses for point-like sources~\cite{Ballmer:2005uw,Mitra:2007mc,Talukder:2010yd}. Despite the different methods used, resulting maps have shown a high level of consistency~\cite{Suresh:2020khz}. Subsequent improvements in map-making techniques have been made, including the implementation of data folding to enhance the speed of map-making significantly~\cite{Ain:2015lea,Ain:2018zvo,Agarwal:2021gvz}.  Phase-coherent mapping have also been developed to map the amplitude and phase parts of the two \ac{GW} polarization modes in the sky~\cite{Cornish:2014rva,Gair:2014rwa,Romano:2015uma,Renzini:2021iim}. Additionally, a spherical harmonic decomposition of the square-root power has been introduced to prevent negative \ac{GW} power in map-making~\cite{Taylor:2020zpk,Payne:2020pmc,Banagiri:2021ovv}. 
Regarding actual analyses, while definitive proof of anisotropy in pulsar timing data is still pending official confirmation, there are indications that the dipole signal may be approaching significant levels based on current data~\cite{NANOGrav:2023tcn}. Collaborations involving LIGO, Virgo, and KAGRA have used cross-correlation data from ground-based detector networks to construct sky maps of \acp{SGWB}, with a focus on compact object mergers. Their findings suggested that the dipole resulting from Earth's peculiar motion is currently well below the detection limits of existing ground-based detectors by more than an order of magnitude~\cite{LIGOScientific:2011kvu,LIGOScientific:2016nwa,KAGRA:2021mth}. 
	
Future space-borne detectors show similarities in \ac{SGWB} detection methods compared to \ac{PTA} and ground-based detectors. Although real data is lacking, studies have performed map-making using generated data, primarily focusing on the individual LISA detector. Taruya et al. reconstructed skymaps of \ac{SGWB} at both low and high frequencies, free from detector noise~\cite{Taruya:2005yf,Taruya:2006kqa}. Their studies emphasized the potential to increase the maximum angular resolution $l_{\rm max}$ for high-frequency skymaps from 5 to 10 in comparison to low-frequency skymaps. Considering the impact of detector noise, $l_{\rm max}$ can be extended up to 15 when clear signals are above the noise level up to $10^{-2}\,\,{\rm Hz}$, but may remain below 7 due to the dominant Galactic foreground at frequencies below $10^{-2}\,\,{\rm Hz}$~\cite{Contaldi:2020rht}. Utilizing the Bayesian spherical harmonic approach, Banagiri et al.~\cite{Banagiri:2021ovv} mapped the Galactic foreground with $l_{\rm max}\leq2$. Apart from LISA, other space-borne detectors are anticipated in the future. The strategic placement of space-borne detectors enables optimal detector separation, promising a significant boost in sensitivity to anisotropy. This advancement is expected to open up new possibilities for high-resolution map-making. The implementation of data analysis for the map-making of \ac{SGWB} in the mHz band is a crucial step that we plan to undertake in the near future.

\begin{acknowledgments}
This work has been supported by the Guangdong Major Project of Basic and Applied Basic Research (Grant No. 2019B030302001), the National Key Research and Development Program of China (No. 2020YFC2201400), the Natural Science Foundation of China (Grants No. 12173104), the Natural Science Foundation of Guangdong Province of China (Grant No. 2022A1515011862), and the Guangdong Basic and Applied Basic Research Foundation(Grant No. 2023A1515030116). 
Z.C.L. is supported by the China Postdoctoral Science Foundation (Grant No. 2023M744094), and the Guangdong Basic and Applied Basic Research Foundation(Grant No. 2023A1515111184). 
We also thank Zu-Cheng Chen, Jianwei Mei, Sai Wang for helpful discussions. 
\end{acknowledgments}

\appendix
\bw
\section{Useful derivation for overlap reduction function}\label{appen:ORF}

To begin, let us first establish the definition of the total ORF for the detector network, assuming the noise \acp{PSD} of the $I$ and $J$ channels within the same detector are equal:
\bea
\label{eq:orf_tot}
\nn
\Gamma_{\rm tot}^{(l)}\left(f,\Delta\vec{x},\{p_{i}\}\right)
&=&\sqrt{\sum_{I={\rm A,E}\atop J={\rm A',E'}}
	\left|\Gamma_{IJ}^{(l)}(f,\Delta\vec{x},\{p_{i}\})\right|^{2}}\\
\nn
&=&\frac{1}{8\pi \sqrt{(2l+1)}}
\sqrt{\sum_{IJ}
\sum_{m=-l}^{l}\left|\sum_{P=+,\times}
\int_{S^{2}}{\rm d}\hat{\Omega}_{\hat{k}}
\mathbb{Y}_{IJ}\left(f,\hat{k},\{p_{i}\}\right)
\frac{Y_{lm}(\hat{k})}{Y_{00}(\hat{k})}
e^{-{\rm i}2\pi f\hat{k}\cdot\Delta\vec{x}/c}\right|^{2}},\\
\eea
where the parameter group $p_{i}$ is related to the geometry of detector network, such as the detector plane angle, the angle between the separation vector and the two detector planes. 
The function 
\be
\mathbb{Y}_{IJ}\left(f,\hat{k},\{p_{i}\}\right)
=F_{I}^{P}\left(f,\hat{k},\{p_{i}\}\right)
F_{J}^{P*}\left(f,\hat{k},\{p_{i}\}\right).
\ee
When $f \ll c/(2 \pi L)$, for two gravitational waves propagating in opposite directions,
\be
\mathbb{Y}_{IJ}\left(f,\hat{k},\{p_{i}\}\right)
=\mathbb{Y}_{IJ}\left(f,-\hat{k},\{p_{i}\}\right).\\
\ee
Since the spherical harmonics
\be
Y_{lm}(\hat{k})=(-1)^{l}Y_{lm}(-\hat{k}),
\ee
By summing the integral terms of \ac{ORF} for two gravitational waves propagating in opposite directions, we have 
\bea
\label{eq:Gamma_IJ}
\nn
&&\mathbb{Y}_{IJ}\left(f,\hat{k},\{p_{i}\}\right)
Y_{lm}(\hat{k})
{\rm e}^{-{\rm i}2\pi f \hat{k}\cdot \Delta \vec{x}/c}+
\mathbb{Y}_{IJ}\left(f,-\hat{k},\{p_{i}\}\right) Y_{lm}(-\hat{k})
{\rm e}^{-{\rm i}2\pi f (-\hat{k})\cdot \Delta \vec{x}/c}\\
\nn
&=&
\left[{\rm e}^{-{\rm i}2\pi f \hat{k}\cdot \Delta \vec{x}/c}
+(-1)^{l}{\rm e}^{{\rm i}2\pi f \hat{k} \cdot \Delta \vec{x}/c}\right]\mathbb{Y}_{IJ}\left(f,\hat{k},\{p_{i}\}\right) Y_{lm}(\hat{k})\\
&=&
\left\{
\begin{array}{lr}
	2\mathbb{Y}_{IJ}\left(f,\hat{k},\{p_{i}\}\right)
	Y_{lm}(\hat{k}) 
	\cos \left[\frac{2\pi f \hat{k} \cdot \Delta \vec{x}}{c}\right],\quad l={\rm even} \\
	\\
	-{\rm i}2\mathbb{Y}_{IJ}\left(f,\hat{k},\{p_{i}\}\right) Y_{lm}(\hat{k})
	\sin \left[\frac{2\pi f \hat{k} \cdot \Delta \vec{x}}{c}\right],\quad l={\rm odd}
\end{array}.
\right.
\eea
The e-index term in the \ac{ORF} leads to a cancellation effect of gravitational waves propagating in the opposite direction, resulting in a significant drop in the ORF near the frequencies $c/(4\Delta |\vec{x}|)$ and $c/(2\Delta |\vec{x}|)$ for even and odd orders, respectively. For a given value of $l$ and $m$, the \ac{ORF} $\Gamma^{lm}_{IJ}$ exhibits a pattern of transitioning from positive to negative values, with its zero point located near the characteristic frequency. 

Next, let us delve deeper into ORF. 
Based on the coordinate rotation invariance of the total ORF, it is possible to align the separation vector $\Delta \vec{x}$ along the $Z$-axis. 
Subsequently, utilizing the spherical wave expansion with the spherical Bessel function $j_{l}$, the exponential term of~\eq{eq:orf_tot} can be expressed as follows:
\bea
\label{eq:e_term}
\nn
{\rm e}^{-{\rm i}2\pi f \hat{k}\cdot \Delta \vec{x}/c}
&=&
4 \pi \sum_{l m}i^{l}Y^{*}_{lm}(-\hat{k})Y_{lm}(\hat{Z})
j_{l}\left(\frac{f}{f_{\ast}}\right)\\
&=&
\sum_{l}\sqrt{4 \pi (2l+1)}i^{l}Y_{l0}(\hat{k})
j_{l}\left(\frac{f}{f_{\ast}}\right),
\eea
where the characteristic frequency $f_{\ast}=c/(2\pi f |\Delta \vec{x}|)$. 
Additionally, we can break down the following function into spherical harmonics:
\be
\frac{1}{8\pi\,Y_{00}(\hat{k})}\sum_{P}
F_{I}^{P}\left(f,\hat{k},\{p_{i}\}\right)
F_{J}^{P*}\left(f,\hat{k},\{p_{i}\}\right)=
\sum_{lm}\mathcal{F}^{IJ}_{lm}\left(f,\{p_{i}\}\right)
Y_{lm}(\hat{k}),
\ee
with the coefficient
\be
\label{eq:FF}
\mathcal{F}^{IJ}_{lm}\left(f,\{p_{i}\}\right)=
\frac{1}{8\pi\,Y_{00}(\hat{k})}\sum_{P}
\int_{S^{2}}{\rm d}\hat{\Omega}_{\hat{k}}
F_{I}^{P}\left(f,\hat{k},\{p_{i}\}\right)
F_{J}^{P*}\left(f,\hat{k},\{p_{i}\}\right)Y^{*}_{lm}(\hat{k}).
\ee
Combined \eq{eq:e_term} with \eq{eq:FF}, we have
\bea
\label{eq:orf_IJ}
\nn
\Gamma_{IJ}^{(l)}(f,\Delta\vec{x},\{p_{i}\})
&=&\frac{1}{\sqrt{(2l+1)}}\sqrt{\sum_{m}\left|
	\int_{S^{2}}{\rm d}\hat{\Omega}_{\hat{k}}
	\sum_{l_{\rm f}m_{\rm f}}
	\mathcal{F}^{IJ}_{l_{\rm f}m_{\rm f}}
	\left(f,\{p_{i}\}\right)
	Y_{l_{\rm f}m_{\rm f}}(\hat{k})Y_{lm}(\hat{k})
	\sum_{l_{\rm x}}
	\sqrt{4 \pi (2l_{\rm x}+1)}{\rm i}^{l_{\rm x}}Y_{l_{\rm x}0}(\hat{k})
	j_{l_{\rm x}}\left(\frac{f}{f_{\ast}}\right)\right|^2}\\
&=&
\sqrt{\sum_{m}\left|\sum_{l_{\rm f}l_{\rm x}} {\rm i}^{l_{\rm x}}
	\sqrt{2l_{\rm f}+1}(2l_{\rm x}+1)
	\Bigg(\begin{matrix}
		l_{\rm f}  & l  &  l_{\rm x} \\
		-m         & m  &  0 \\
	\end{matrix}
	\Bigg)
	\Bigg(\begin{matrix}
		l_{\rm f}  & l  &  l_{\rm x} \\
		0          & 0  &  0 \\
	\end{matrix}
	\Bigg)
	\mathcal{F}^{IJ}_{l_{\rm f}-m}\left(f,\{p_{i}\}\right)
	j_{l_{\rm x}}\left(\frac{f}{f_{\ast}}\right)\right|^{2}},
\eea
where the integral of a product of spin-weighted spherical harmonics can be obtained by multiplying two Wigner-3j symbols~\cite{Messiah1981QuantumMV,1988qtam.book.....V,Romano:2016dpx}:
\be
\label{eq:inter_sp}
\int_{S^{2}}{\rm d}\hat{\Omega}_{\hat{k}}\,\,\,
_{s_{1}}Y_{l_{1}m_{1}}(\hat{k})\,_{s_{2}}Y_{l_{2}m_{2}}(\hat{k})\,_{s_{3}}Y_{l_{3}m_{3}}(\hat{k})=
\sqrt{\frac{(2 l_{1}+1) (2 l_{2}+1) (2 l_{3}+1)}{4 \pi}}
\Bigg(\begin{matrix}
	l_{1}  & l_{2}  &  l_{3} \\
	m_{1}  & m_{2}  &  m_{3} \\
\end{matrix}
\Bigg)
\Bigg(\begin{matrix}
	l_{1}  & l_{2}  &  l_{3} \\
	-s_{1}  &-s_{2}  & -s_{3} \\
\end{matrix}
\Bigg),
\ee
with the following conditions that ensure the Wigner-3j symbols are not equal to 0:
\bea
\label{eq:W3_cond}
\begin{array}{lr}
	|l_{2}-l_{1}|\leq l_{3} \leq |l+l_{1}| \\	
	l_{1}+l_{2}+l_{3}={\rm even} \\
	m_{1}+m_{2}+m_{3}=0
\end{array}.
\eea
Furthermore, when $f \ll c/(2\pi L)$ with the detector armlength $L$, the non-vanishing terms of $\mathcal{F}^{IJ}_{l_{\rm f}m_{\rm f}}$ are given by $l_{\rm f}=0,2,4$, and, as such,
\bea
\label{eq:l_x}
l_{\rm x}=
\left\{
\begin{array}{lr}
	0,2,...,l+4,&l=0,2,4\\	
	1,3,...,l+4,&l=1,3,5\\
	l-4,l-2,...,l+2,l+4,& l\geq 6
\end{array}.
\right.
\eea
Then according to~\eq{eq:orf_IJ}, one can calculate the zero-valued solution of ORF. 
For $l=0$, the total ORF is solely contributed by harmonics with $m=0$:
\bea
\label{eq:ORF_tot_l0}
\nn
\Gamma_{\rm tot}^{00}(f,\Delta\vec{x},\{p_{i}\})&=&
\frac{3\sqrt{\pi}}{70}
\sin\left[2\pi fL/c\right]\sin\left[2\pi fL'/c\right]
\left[42\,j_{0}\left(\frac{f}{f_{\ast}}\right) - 12\sqrt{5}\,j_{2}\left(\frac{f}{f_{\ast}}\right) + 
j_{4}\left(\frac{f}{f_{\ast}}\right)\right],\\
\eea
where the two detectors are positioned in the $Y$-$Z$ plane and the $X$-$Y$ plane. 
Within the frequency range where $f/f_{\ast}$ is less than 10, the zero-valued solutions for $f/f_{\ast}$ within \eq{eq:ORF_tot_l0} are \{2.66,6.07,9.29\}. 
Consequently, the total ORF for $l=0$ can reach zero near the characteristic frequency. 
Regarding $l=1$, it is noteworthy that $|\Gamma_{\rm tot}^{lm}|=|\Gamma_{\rm tot}^{l-m}|$, allowing us to concentrate on the harmonics with $m=0,1$: 
\bea
\label{eq:ORF_tot_l1}
\nn
\Gamma_{\rm tot}^{10}(f,\Delta\vec{x},\{p_{i}\})&=&
\frac{\sqrt{\pi}}{70}
\sin\left[2\pi fL/c\right]\sin\left[2\pi fL'/c\right]
\left[198\,j_{1}\left(\frac{f}{f_{\ast}}\right) - 112\,j_{3}\left(\frac{f}{f_{\ast}}\right) + 
5\,j_{5}\left(\frac{f}{f_{\ast}}\right)\right],\\
\Gamma_{\rm tot}^{11}(f,\Delta\vec{x},\{p_{i}\})&=&
\frac{\sqrt{\pi}}{35}
\sin\left[2\pi fL/c\right]\sin\left[2\pi fL'/c\right]
\left[54\,j_{1}\left(\frac{f}{f_{\ast}}\right) + 49\,j_{3}\left(\frac{f}{f_{\ast}}\right) - 5\,j_{5}\left(\frac{f}{f_{\ast}}\right)\right].
\eea
The corresponding zero-valued solutions for $f/f_{\ast}$ within \eq{eq:ORF_tot_l1} are \{3.96,7.45\} and 
\{5.57,8.98\}, respectively. 
For $l=1$, the total ORF is contributed by harmonics with $m=0,\pm 1$, and therefore it can no longer drops to zero near the characteristic frequency. 
This conclusion also holds true for higher-order ORFs. 
In addition to analyzing the zero value point of the ORF, let us further explore its properties at the low-frequency and high-frequency limits. 

In the low-frequency limit, the spherical Bessel function~\cite{ARFKEN2013643}
\be
\label{eq:jl_low}
j_{l}\left(\frac{f}{f_{\ast}}\right) \simeq 
\frac{1}{(2l+1)!!}\left(\frac{f}{f_{\ast}}\right)^l
\propto |\Delta \vec{x}|^l.
\ee
Here, the symbol $!!$ denotes the double factorial. Then in terms of \eq{eq:orf_tot}, \eq{eq:orf_IJ}, \eq{eq:l_x} and \eq{eq:jl_low}, when $f\ll c/(2\pi |\Delta \vec{x}|)$, the total ORF
\bea
\Gamma_{\rm tot}^{(l)}(f)\propto\left\{
\begin{array}{lr}
	f^{2},&l=0,2,4 \\
	\\
	\frac{|\Delta \vec{x}|}{c}\,f^{3},&l=1,3,5\\
	\\
	\frac{|\Delta \vec{x}|^{l-4}}{c}\,f^{l-2},& l\geq 6
\end{array},\quad f\ll c/(2\pi|\Delta \vec{x}|),
\right.
\eea
where $f^{2}$ derived from the normalized factor $C_{\rm TDI}$. \eq{eq:orf_IJ} and \eq{eq:l_x} indicate that as the order $l$ of the ORF increases, the order of the spherical Bessel functions involved in its calculation also increases. 
This higher order corresponds to larger first non-zero roots of the spherical Bessel function. 
However, it is crucial to note that the above conclusion may not hold true when the dropping frequency exceeds $c/(2 \pi L)$. 

\begin{figure*}[htbp]
	\centering
	\includegraphics[height=6cm]{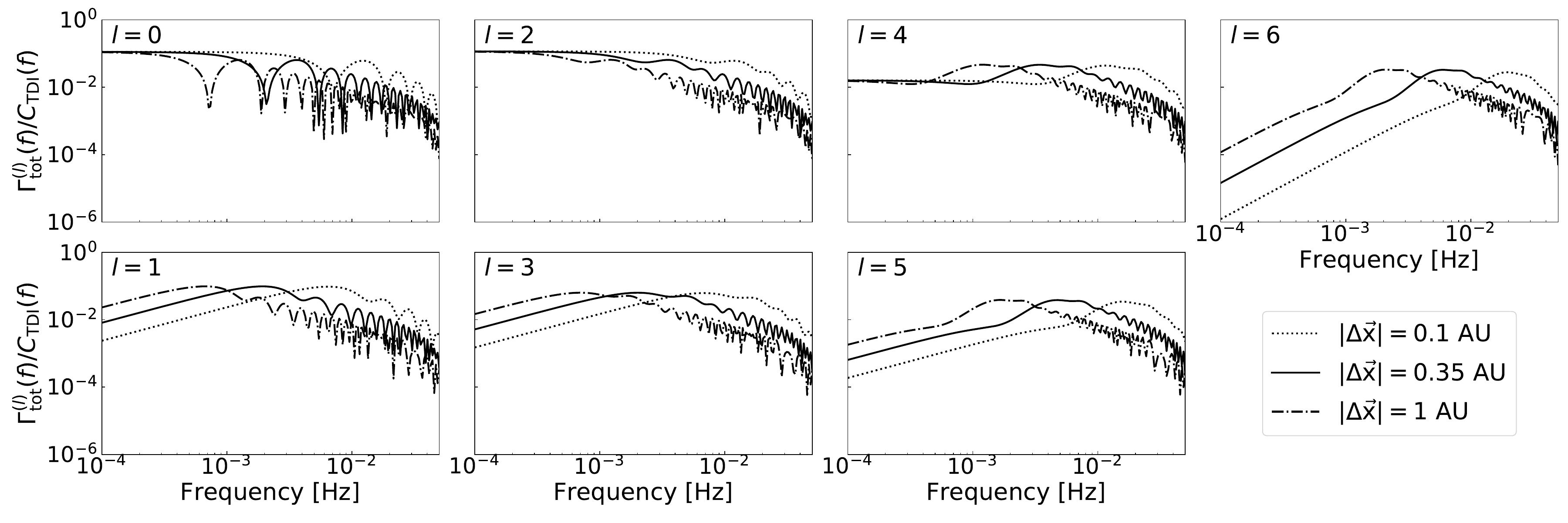}
	\caption{ORF of the TianQin + LISA network with different detector separations.}
	\label{fig:ORF_tl_l_dx}
\end{figure*}

In the high-frequency limit, the spherical Bessel function~\cite{ARFKEN2013643}
\be
j_{l}\left(x\right) \simeq 
\frac{1}{x}\sin\left(x-l \pi/2\right).
\ee
Then one can define 
\be
\label{eq:J_l}
\mathcal{J}_{l}(x)
={\rm i}^{l}\frac{1}{x}\sin\left(x-l \pi/2\right)
=\left\{
\begin{array}{lr}
	\frac{1}{x}\sin x,l={\rm even}\\	
	\\
	\frac{{\rm i} }{x}\cos x,l={\rm odd}
\end{array},
\right.
\ee
of which the amplitude is independent of the order $l$.
Furthermore, through the orthogonality relation of Wigner 3j symbols~\cite{Messiah1981QuantumMV}
\be
\sum_{lm}(2l+1)
\Bigg(\begin{matrix}
	l_{1}  & l_{2}  &  l \\
	m_{1}  & m_{2}  &  m \\
\end{matrix}
\Bigg)
\Bigg(\begin{matrix}
	l_{1}  & l_{2}  &  l \\
	m'_{1}  & m'_{2}  &  m \\
\end{matrix}
\Bigg)
=\delta_{m_{1}m'_{1}}\delta_{m_{2}m'_{2}},
\ee
we have 
\be
\label{eq:W3_ort}
\sum_{l_{\rm x}m_{\rm x}}(2l_{\rm x}+1)
\Bigg(\begin{matrix}
	l_{\rm f}  & l  &  l_{\rm x} \\
	-m         & m  &  m_{\rm x} \\
\end{matrix}
\Bigg)
\Bigg(\begin{matrix}
	l_{\rm f}  & l  &  l_{\rm x} \\
	0          & 0  &  m_{\rm x} \\
\end{matrix}
\Bigg)
=
\sum_{l_{\rm x}}(2l_{\rm x}+1)
\Bigg(\begin{matrix}
	l_{\rm f}  & l  &  l_{\rm x} \\
	-m         & m  &  0 \\
\end{matrix}
\Bigg)
\Bigg(\begin{matrix}
	l_{\rm f}  & l  &  l_{\rm x} \\
	0          & 0  &  0 \\
\end{matrix}
\Bigg)
=
\delta_{m0},
\ee
where $\delta_{ab}$ is the Kronecker delta function, and the third line of \eq{eq:W3_cond} is adopted. 
To satisfy the condition for the second Wigner-3j symbol in \eq{eq:W3_ort} to be non-zero, which is $l_{\rm f} + l + l_{\rm x}={\rm even}$~\cite{Messiah1981QuantumMV}, it is necessary to have $l_{x}=|l-l_{f}|,|l-l_{f}|+2,...,|l+l_{f}|-2,|l+l_{f}|$. 
By following the above rule and employing \eq{eq:J_l} and \eq{eq:W3_ort}, \eq{eq:orf_IJ} can be simplified to
\be
\label{eq:ORF_IJ}
\Gamma_{IJ}^{(l)}(f,\Delta\vec{x},\{p_{i}\})=
\sqrt{\left|\sum_{l_{\rm f}}\sqrt{2l_{\rm f}+1}
	\mathcal{F}^{IJ}_{l_{\rm f}0}\left(f,\{p_{i}\}\right)
	\mathcal{J}_{l_{\rm f}+l}
	\left(\frac{f}{f_{\ast}}\right)\right|^{2}}
\propto \frac{1}{|\Delta \vec{x}|},
\ee
which implies that in the high-frequency limit, the total ORF is inversely proportional to the detector separation and remains independent of the order $l$. Taking the TianQin + LISA network as an example, \fig{fig:ORF_tl_l_dx} shows that the \ac{ORF} decreases with the increase of detector separation in the high-frequency limit. 

\ew

\normalem
\bibliographystyle{apsrev4-1}
\bibliography{Paper_II}

\end{document}